\documentclass[manuscript]{emulateapj}
\usepackage{epsfig}
\usepackage{apjfonts}
\usepackage{epsfig}
\usepackage{amsmath}
\usepackage{longtable}

\newcommand{\lsim}{\lower0.6ex\vbox{\hbox{$ \buildrel{\textstyle
        <}\over{\sim}\ $}}}
\newcommand{\gsim}{\lower0.6ex\vbox{\hbox{$ \buildrel{\textstyle
        >}\over{\sim}\ $}}}

\newcommand{\beq}{\begin{equation}}
\newcommand{\eeq}{\end{equation}}

\shorttitle{SMBH Mass - Pitch Angle Relation}
\shortauthors{Berrier et~al.}

\submitted{Published in \apj, 769, 132}

\begin{document}

\title{Further Evidence for a Supermassive Black Hole Mass - Pitch Angle Relation
}

\author{Joel C.\ Berrier \altaffilmark{1,2,3,4}}

\author{Benjamin L.\ Davis \altaffilmark{2}}

\author{Daniel Kennefick \altaffilmark{1,2}}

\author{Julia D. Kennefick \altaffilmark{1,2}}

\author{Marc S.\ Seigar \altaffilmark{2,5}}   

\author{Robert Scott Barrows \altaffilmark{2}}

\author{Matthew Hartley \altaffilmark{1}}

\author{Doug Shields \altaffilmark{1,2}}

\author{Misty C. Bentz \altaffilmark{6}}

\author{Claud H.~S.\ Lacy \altaffilmark{1,2}}

\altaffiltext{1}{Department of Physics, University of Arkansas, 825 West Dickson
  Street, Fayetteville,  AR 72701, USA}

\altaffiltext{2}{Arkansas Center for  Space and
  Planetary Sciences, University of Arkansas,
202  Old Field House, Fayetteville, AR 72701, USA}

\altaffiltext{3}{Instituto de Astrofisica de Canarias,
C/ Via Lactea s/n, E-38200 Tenerife, Spain}

\altaffiltext{4}{Dept. Astrofisica,Universidad de La Laguna, E-38206 La Laguna, Tenerife, Spain }

\altaffiltext{5}{Department  of  Physics  and  Astronomy,
  University of Arkansas at Little Rock, 2801 South University Avenue,
  Little Rock,  AR 72204, USA} 

\altaffiltext{6}{Department of Physics and Astronomy, Georgia State University,
Atlanta, GA 30303, USA}

\begin{abstract}
We present new and stronger evidence for a previously reported
relationship  between galactic
spiral  arm pitch  angle $P$  (a measure  of the  tightness of spiral
structure)  and the  mass  $M_{\rm  BH}$ of  a  disk galaxy's nuclear
supermassive  black  hole  (SMBH).   We  use  an  improved  method to
accurately  measure the  spiral arm  pitch angle  in disk  galaxies to
generate  quantitative data  on  this morphological  feature for $34$
galaxies with directly measured black hole masses.  We find a relation
of $\log(M/M_{\odot}) = (8.21\pm0.16) - (0.062\pm0.009)P$.  This method
is  compared  with  other  means  of estimating  black  hole  mass to
determine its effectiveness and  usefulness relative to other existing
relations. We argue that such a relationship is predicted
by leading theories of spiral structure in disk galaxies, including
the density wave theory. We propose this  relationship as a
tool  for  estimating SMBH  masses  in  disk  galaxies. This  tool is
potentially superior when compared to  other methods for this class of
galaxy and has the advantage of being unambiguously measurable from
imaging data alone.
\end{abstract}



\section{Introduction} \label{sec:intro}

Since the existence  of supermassive black holes (SMBHs)  as a common,
or even ubiquitous, component of  galactic bulges was first recognized
\citep{Kormendy&Richstone1995,         Barth2004,        Kormendy2004,
  Magorrian1998}, increasingly  successful attempts have been  made to
measure the  mass of  these objects. This  has enabled  astronomers to
discover correlations between the mass of an SMBH and its host galaxy's
mass   or   luminosity  \citep{Kormendy1993,   Kormendy&Richstone1995,
  Magorrian1998, Marconi2003, Haring2004}. A number of features of the
host galaxy have now been found to  correlate to the mass of the black
hole, giving rise to efforts to study the black holes by making
measurements of features of the host  galaxy even where the black hole
is undetectable.  Although  many of these correlating  features of the
host galaxy require spectroscopy to measure, one which does not is the
S\'ersic  index  of  the   galaxy's  bulge  \citep{Graham2007}.   This
correlation  demonstrates the  feasibility of  estimating SMBH  masses
through  imaging data  alone.  Here  we  verify and  further refine  a
recently discovered relation  between the spiral arm pitch  angle of a
galaxy and the mass of its SMBH, the $M$--$P$ relation \citep{SKKL08}.

Our knowledge  of SMBH masses  in the universe has  grown dramatically
over the  last decade, primarily due to  high-resolution observations
made with  the {\it Hubble Space  Telescope (HST)}. These  observations have
shown that SMBHs  reside not only in the cores  of active galaxies, as
has been  believed for decades, but  also in the centers  of quiescent
galaxies.  Recent  works have begun  to explore the importance  of the
nuclear SMBHs  in the evolution,  or co-evolution, of its  host galaxy
\citep[e.g.][]{Magorrian1998,   Ferrarese&Merritt2000,  Gebhardt2000a,
  Marconi2003,  Haring2004,  Springel2005,  Hopkins2007,  Rosario2010,
  Crenshaw2010, Treuthardt2012}.  As a  result, any complete theory of
galaxy  formation has  to  produce  SMBHs in  the  centers of  massive
galaxies \citep[e.g.][]{Silk&Rees1998},  and explain the  evolution of
SMBH mass over time.

Direct determination of SMBH mass depends on instrumentation which can
observe the motion  of stars, gas, and dust in  the immediate vicinity
of  the   black  hole   \citep{Kormendy&Richstone1995,  Macchetto1997,
  Maciejewski2001}.   This process  is  observationally expensive  or
impossible for distant galaxies.  It  is natural, therefore, that much
time and effort have gone into the search for an indirect measure which
can be  used to give estimates  of central SMBH mass.   One remarkable
indicator  is the  $M_{\rm BH}$--$\sigma$  relation, which  relates the
central SMBH  mass ($M_{\rm  BH}$) to the  velocity dispersion  in the
central   galactic   bulge  ($\sigma$)   \citep{Ferrarese&Merritt2000,
  Gebhardt2000a}.   The  $M_{\rm BH}$--$\sigma$   relation  has  led  to
considerable  success  estimating  the  SMBH mass  for  somewhat  more
distant galaxies, generally those whose cosmological redshift ($z$) is
$z <  0.1$ \citep{Heckman2004, Kauffmann2007}.   Thus the mass  of the
SMBH can be estimated for a wider range of galaxies.  Nevertheless, as
it  is  necessary  to  have  spectroscopic  measurements  to  estimate
$\sigma$ for the spheroidal component,  it is still expensive in terms
of telescope  time.  Additionally, measuring $\sigma$  is more complex
for disk galaxies  than it is for ellipticals, since  one must account
for velocity  dispersion associated  with the motion  of disk  and bar
stars intermingled  with the velocity  dispersion of the  bulge stars,
which is actually the  correlating quantity \citep{Hu2008}. Studies do
suggest that the scatter in the $M_{\rm BH}$--$\sigma$ is significantly
greater for  disk galaxies than for  ellipticals \citep{Gultekin2009}.
Although  projects such  as the  Sloan  Digital Sky  Survey have  made
progress in acquiring spectra of large numbers of galaxies, there have
been few ways, up to now, to take advantage of the far larger catalogs
of imaging data available from the  public archives for the purpose of
estimating black hole masses.

A far  greater number of  estimates can be  made if features  that are
linked  to the  mass  of  SMBHs can  be  measured  from imaging  data.
Several  such relations  have been  explored, including  those between
black        hole       mass        and       bulge        luminosity
\citep[e.g.,][]{Kormendy&Richstone1995,Haring2004},  and  S\'ersic  index
and  nuclear  SMBH  mass,   \citep{Graham2007},  among  others.   More
recently, a relation between the central  SMBH mass and the spiral arm
pitch angle of  its host galaxy has been discovered  by examination of
$27$   disk   galaxies   with   previously   estimated   SMBH   masses
\citep{SKKL08}.   The pitch  angle of  the spiral  arms of  the galaxy
(essentially how tightly  the spiral structure in the  arms are wound)
can be measured solely from images of the galaxy.  There exist sets of
images which  cover significant field-of-view (FOV)  to a considerable
depth  and  look-back time  for  which  this technique  could  provide
estimates of the SMBH mass in  a complete sample of spiral galaxies at
much  greater  distances  than  have  been  possible  hitherto,  up  to
redshifts $z \sim 1$ for especially deep images and favorable objects,
but very likely to a redshift of $z \sim 0.5$ for a significant sample
of galaxies.

We  may ask  ourselves if  such a  relationship is  expected from  our
understanding of galactic physics.  The  answer is in the affirmative.
First, it  has been shown empirically  that there is a  link between
spiral arm pitch angle and  the central mass concentration of galaxies
\citep{Seigar2005,  Seigar2006}.   It  is   generally  agreed  that  a
``strong  correlation between  central  mass  concentration and  pitch
angle [is] predicted by modal density wave theory,'' \citep{Grand2012}
the  dominant theory  of  spiral arm  structure.   Furthermore such  a
correlation between the size of the central bulge and the tightness of
the spiral arm winding  constitutes essentially the first significant,
though at that point largely qualitative, observation of extragalactic
astronomy, the  basis of  the Hubble  classification. The  notion that
black hole mass  depends on the mass of the  central galactic bulge is
now also widely  established, as a result of  observed correlations of
black  hole  mass  with  both  bulge  velocity  dispersion  and  bulge
luminosity.  Therefore black  hole  mass and  spiral  arm pitch  angle
should  each  measure  the   central  mass  concentration  and  should
therefore  correlate with  each  other quite  strongly.  Although  the
mechanism by  which the  correlation of black  hole mass  with central
bulge mass  is maintained is  uncertain there are  nevertheless highly
plausible      arguments      why      they      should      correlate
\citep[e.g.,][]{Silk&Rees1998}.

\citet{Shu1984}, who deals primarily with  the case of density waves
in the rings  of Saturn naturally focuses on what  can, in the context
of galactic astronomy,  be termed the bulge-dominated case.  One has a
central body (Saturn) whose mass far outweighs that of the material in
the  disk.  The result  is  relatively  straightforward to  deal  with
theoretically, producing density waves in a tractable short wavelength
approximation (density  waves in  Saturn's rings typically  have pitch
angles measured in tenths of a degree). Even better the system is well
understood observationally  and there  is excellent  agreement between
theory and  observation. We  may thus  be very  happy with  the result
given by \citet{Shu1984} that the tangent  of the pitch angle ($i$) of
the spiral pattern produced by density waves should be proportional to
the ratio of the surface density of  material in the disk to the total
mass of the  central body. Although \citet{Shu1984}  focuses mostly on
the case of Saturn the formalism (the theory was, of course, developed
in the context  of galactic spiral arms) is  still approximately valid
for  the most  bulge-dominated disk  galaxies with  relatively tightly
wound spirals (small $i$ and thus with the shortest wavelength density
waves seen in galactic disks).

Naturally the  case of  galactic disks is  much more  complicated than
that of Saturn's  rings, not least because of the  self-gravity of the
disk itself. Nor can we say that there is such close agreement between
theory and observation in this  case. Nevertheless there is quite good
agreement and  some success has  been achieved in  modeling individual
galaxies   with    the   density   wave   theory.    An   example   is
\citet{Roberts1975} which shows that the  pitch angle in disk galaxies
depends on  the ratio of two  radii, the half-mass radius,  defined as
the  radius  within which  half  the  mass  of  the galaxy's  disk  is
contained, and the corotation radius,  defined as the radius at which
stars and other material bodies in the disk rotate at the same rate as
the  spiral  pattern.   Thus,  once  again,  we  see   that  the  more
concentrated the  mass of the galaxy  is toward the center  (and thus
the smaller is  the half-mass radius), the tighter will  be the spiral
pattern.

It  is not  hard to  show that  the result  of \citet{Roberts1975}  is
compatible with  \citet{Shu1984}. If  one reduces the  former's Toomre
disk model  to a very  simple bulge (or planet)  with a thin  low-mass
disk  of uniform  density  and thickness,  then  the half-mass  radius
shrinks (and thus the pitch angle decreases) depending on the ratio of
the central mass (the planet or  bulge) to the surface mass density of
the disk, which is the  controlling factor in \citet{Shu1984}. Thus in
bulge-dominated galaxies the pitch angle depends inversely on the mass
of the central bulge. Disk-dominated galaxies, especially the extreme
case of  bulgeless galaxies,  behave similarly   in that  their pitch
angle correlates  to the  relative concentration  of mass  toward the
center of the galaxy. Unfortunately,  relatively little is known as yet
about the relation between black  hole mass and galaxy characteristics
in  disk-dominated galaxies,  since relatively  few black  hole masses
have been directly measured in such galaxies.

Indeed if there is no classical bulge,  it is difficult to know how to
interpret   the  $M$--$\sigma$   or  $M$--bulge   luminosity  relations   at
all. Clearly,  further work  will have  to be  done to  understand the
relation between disk dominated galaxies and their central black holes
(if they  have one). However,  there are arguments which  might suggest that there should still be  a link between black hole mass  and the mass of
the central part  of the disk. It is widely  suspected that bulges are
produced by  mergers in  which the central  parts of  galaxies become
hotter and, heated  past the point in which they  maintain a flattened
disk profile,  adopt a bulge profile.  If the central black  hole mass
does indeed  correlate to the  mass of this merger-created  bulge, one
might speculate  that it  would have  previously (before  the mergers)
correlated to the mass of the central region of the disk, out of which
material the  post-merger bulge  was presumably  formed.  But  in that
case the spiral arm pitch angle, pre-merger, would have also tended to
correlate to  the mass of  the disk's  central region (which  tends to
control  the  value of  the  half-mass  radius).  Although it  is  not
possible to  say anything with certainty  at this stage, it  may prove
that the mass--pitch  angle relation could work  for disk-dominated and
bulgeless galaxies  where other  correlations ($M$--$\sigma$,  $M--L$) would
need to be reinterpreted. In the  meantime, we can be fairly confident
that for galaxies with classical bulges  the pitch angle of the spiral
arms should correlate well to the mass of that central bulge.

The dependence  of pitch angle  on central  mass in the  modal density
wave theory  can be  understood by  analogy with  standing waves  on a
string,  since  the  modal   density  waves  themselves  constitute  a
standing-wave pattern.  In  the case of waves on a  string, one expects
the wavelength of the waves oscillating between the ends of the string
to  depend on  the speed  of  propagation of  the wave.  This in  turn
depends  on the  ratio $\rho/T$  where $\rho$  is the  density of  the
string  and $T$  is the  tension in  the string,  the restoring  force
producing the  wave phenomenon.   In the modal  theory, the  density of
material in the  disk $\sigma_o$ plays the role of  the density of the
string  and  the  restoring  force  or  tension  is  produced  by  the
gravitational field of  the massive central region of  the galaxy.  It
is natural  that the  wavelength of the  resultant spiral  pattern should
depend on the ratio of these two quantities.

It should be noted that the modal  density wave theory is not the only
theory  which  attempts  to  explain  spiral  arm  structure  in  disk
galaxies.   Rivals include  the swing  amplification model  of density
wave  theory   \citep{Toomre1981,  Gerola1978,  Seiden1979}   and  the
Manifold theory (see below for  selected references). It would be fair
to say that the modal density  wave theory is the most widely accepted
but that each  of these has significant support, at  least for certain
types of  spiral galaxies. It  has even been suggested  that different
galaxies (for instance, grand design versus flocculant) have different
mechanisms  explaining their  spiral  structure. Apart  from a  theory
which  proposes that  spiral arms  are the  result of  stochastic star
formation operated  upon by  differential rotation all  theories agree
that there is a link between  central mass and spiral arm pitch angle.
The Toomre density wave theory differs from the modal theory primarily
in denying that spiral arm structure lasts for longer than about a few
rotational periods in a given galaxy  (because in this theory there is
no longstanding standing  wave pattern). Pitch angle  should vary with
time in this theory, but should still obey a relation with the size of
the central mass.

The Manifold theory of spiral  structure is the most recently proposed
of      these     theories      \citep{Kaufmann1996,     Harsoula2009,
  Athanassoula2009A,        Athanassoula2009B,       Athanassoula2010,
  Athanassoula2012}.   This theory  describes  the  spiral pattern  as
being  the result  of stars  formed near  the ends  of a  galaxy's bar
moving into chaotic, highly  eccentric orbits which nevertheless cause
the stars  to move along  relatively narrow tubes known  as manifolds.
The  global pattern  produced by  their motion  along these  manifolds
gives rise to the observed spiral arms. The details of this theory are
also subtle, but it is abundantly clear that the orbits, and therefore
the manifolds,  are controlled by  the central mass  concentration, as
with all  galactic orbits, and  that therefore, once again,  the pitch
angle of  the spiral  arms should  vary with the  central mass  of the
galaxy (E. Athanassoula 2012, private communication).

Thus,  we see  that the  primary theory  for the  formation of  spiral
structures, along with  its two main competitors, are  agreed that the
mass of  a central black  hole should correlate  with the mass  of the
central core of the galaxy. These theories demand that the mass of the
central bulge should determine the  pitch angle of the galaxy's spiral
arms. Indeed, it is currently difficult  to imagine a theory of spiral
arm structure which does not demand a correlation with the central mass
concentration,  at  least  indirectly.  But  the  two  currently  most
actively pursued theories (modal density waves and manifold) both give
the  central  mass a  controlling  influence  on the  mechanism  which
generates the  pitch angle of  the spiral pattern.   It is not  at all
surprising,  then, that  we should  find  strong evidence  for such  a
concentration in actual observations, and with a notably low degree of
scatter.

Finally, one needs  look no further than the Hubble  sequence to see an
illustration of the connection  between galactic morphologies and SMBH
mass.   The SMBH  mass-bulge  mass relation,  when  combined with  the
general pattern of larger bulges and  tighter spiral arms as one moves
from Sc to Sa in the Hubble sequence, demonstrates, at the very least,
an indirect  connection between  these properties.   Our view  is that
spiral arm pitch  angle, which appears to be well  correlated at least
with SMBH  mass, would be  an excellent tool  to probe the  complex of
correlated   characteristics   of    spiral   galaxies   for   several
reasons. First, because  it is measurable through  imaging data alone.
Second,  it can  take advantage  of the  great storehouse  of publicly
accessible  archival  data  available.  Finally,  its  measurement  is
independent of redshift, since logarithmic spirals remain self-similar
no matter how they are scaled.  In short, it may be possible that, for
disk galaxies, we can gain information  on the black hole masses for a
significant number of spiral galaxies  which previously could not have
their masses estimated by other means.

Over the  last few decades, it  has become widely accepted  that SMBHs
and  dark  matter  play  influential,  even  dominant,  roles  in  the
evolution of  galaxies.  As  neither black holes  nor dark  matter are
directly  observable  in any  part  of  the electromagnetic  spectrum,
information   about  them   has  been   painstakingly  obtained   from
observations  of  their  gravitational   effect  on  baryonic  matter.
Admittedly, \citet{Kormendy2011A}  suggested that  SMBH mass  does not
correlate  with   galaxy  disks,  and   \citet{Kormendy2011B}  further
suggest that  a galaxy's dark  matter halo  does not have  any direct
correlation  with  the  properties  of  the  nuclear  SMBH.   However,
\citet{Volonteri2011}  and  \citet{Booth:Schaye:2010,Booth:Schaye:2011} all provide counter
results to those of \citet{Kormendy2011A} and \citet{Kormendy2011B}.

In this  paper, we will  re-examine the relationship  of \citet{SKKL08}
and  expand  upon  the  sample  used  in  that  study  by  adding  new
measurements  from other  nearby spiral  galaxies and  active galactic
nuclei  (AGN).  We  will  double  the number  of  points  used in  the
previous work, as well as update the method used in the measurement of
the   spiral  arm   pitch  angles   in  \citet{SKKL08}   to  that   of
\citet{Davis2012}.  We take advantage of SMBH mass data from a variety
of measurement techniques including  direct measurement of stellar and
gas dynamics in the vicinity of  the black hole, measurements based on
available maser data for several objects and reverberation mapping. We
also  make  use of  a  select  set of  data  based  on the  $M$--$\sigma$
relation, in  an effort to deepen  our understanding of the  extent to
which spiral arm structure correlates to central mass.

The    structure    of    this     paper    is    as    follows:    in
Section~\ref{sec:methods},  we outline  the data  we use  in this  work
(including  the   observations  and   mass  determinations)   and  the
techniques  used to  measure  morphological features  of the  observed
galaxies.   In Section~\ref{sec:results},  we assemble  an updated  SMBH
mass--pitch angle relation using a variety of observational results and
compare them across different subsamples.   We also examine the use of
S\'ersic  index as  a  means  of estimating  SMBH  mass from  galactic
morphologies   and  compare   that  method   with  our   results.   In
Section~\ref{sec:Discussion}, we discuss the  implications of this work
and the usefulness of this method of SMBH mass estimation. Finally, in
Section~\ref{sec:Conclusions} we  outline our final assessment  of our
results.

In   this  work,   where   necessary,  we   assume   a  cosmology   of
$\Omega_{\Lambda}  = 0.728$,  $\Omega_b  =  0.0455$, $\Omega_{m}h^2  =
0.1347$, and $H_0 = 70.4$ km s$^{-1}$ Mpc$^{-1}$.  This corresponds to
the maximum likelihood cosmology from the combined {\it WMAP}+BAO+H0 results
from the {\it WMAP} 7 data release \citep{Komatsu2011}.


\section{Methods}
\label{sec:methods}

In  this work,  we investigate  the relation  between spiral  arm pitch
angle  and central  black hole  mass  using galaxies  selected from  a
variety  of sources  with  directly measured  SMBH masses.   Candidate
galaxy  images are  selected from  the available  archival data.   The
sample listed in Table  \ref{table:data}
\begin{deluxetable*}{lrcccccc}
\tablecolumns{8}
\tablecaption{Sample Information}
\tablehead{
\colhead{Galaxy} & \colhead{$P$ (deg.) } & \colhead{Image Source} & \colhead{Filter} & \colhead{$\log$($M_{\rm BH}$/$M_{\odot}$)} &
\colhead{ Measurement Type} & \colhead{Source} & \colhead{Preferred} 
}
\startdata
 3c120     & $ 10.7 \pm 1.4 $ & Danish 1.54 (NED$^a$)                        & $R$                &$7.72^{\tiny{+0.23}}_{\tiny{-0.23}}$ & Reverberation Mapping & 1,2 & Y \\ 
 Ark 120   & $  5.4 \pm 0.6 $ & {\it HST} ACS (Bentz$^b$)                          & F550M            &$8.15^{\tiny{+0.11}}_{\tiny{-0.11}}$ & Reverberation Mapping & 1   & Y \\ 
 Circinus  & $ 26.7 \pm 5.0 $ & {\it HST} WFPC2 (NED$^a$)                          & F814W            &$6.24^{\tiny{+0.07}}_{\tiny{-0.08}}$ & Maser Modeling        & 3   & Y \\ 
 IC 342    & $ 23.2 \pm 2.8 $ & VLA (NED$^a$)                                & 21 cm            &$<5.70$                              & Stars/Gas             & 4   &   \\ 
           &                  &                                              &                  &$6.32^{\tiny{+0.08}}_{\tiny{-0.09}}$ & $M$--$\sigma$            & 5   & Y \\ 
 IC 2560   & $ 16.3 \pm 6.4 $ & 2.5 m du Pont (CGS$^c$)                      & $B$                &$6.64^{\tiny{+0.30}}_{\tiny{-0.30}}$ & Maser Modeling        & 6   & Y \\ 
 M 31      & $  8.5 \pm 1.3 $ & {\it GALEX} (NED$^a$)                              & NUV              &$8.15^{\tiny{+0.22}}_{\tiny{-0.10}}$ & Stars/Gas             & 7   & Y \\ 
           &                  &                                              &                  &$7.59^{\tiny{+0.08}}_{\tiny{-0.10}}$ & $M$--$\sigma$            & 5   &   \\ 
 M 33      & $ 34.5 \pm 8.6 $ & {\it Spitzer} IRAC (NED$^a$)                       & IRAC 3.6 $\mu$m  &$<3.48$                              & Stars/Gas             & 8   &   \\ 
           &                  &                                              &                  &$4.24^{\tiny{+0.08}}_{\tiny{-0.09}}$ & $M$--$\sigma$            & 5   & Y \\
 Mrk 590   & $  8.5 \pm 3.3 $ & {\it HST} ACS (Bentz$^b$)                          & F550M            &$7.66^{\tiny{+0.12}}_{\tiny{-0.12}}$ & Reverberation Mapping & 1   & Y \\ 
 Mrk  79   & $ 13.2 \pm 3.6 $ & Lick 1m (NED$^a$)                            & $V$                &$7.70^{\tiny{+0.16}}_{\tiny{-0.16}}$ & Reverberation Mapping & 1   & Y \\ 
 Mrk 817   & $  9.9 \pm 4.2 $ & {\it HST} ACS (Bentz$^b$)                          & F550M            &$7.69^{\tiny{+0.08}}_{\tiny{-0.07}}$ & Reverberation Mapping & 2   & Y \\
 Milky Way & $ 22.5 \pm 2.5 $ & Leiden/Argentine/Bonn (LAB) Survey (NED$^a$) & 21 cm            &$6.63^{\tiny{+0.04}}_{\tiny{-0.04}}$ & Stars/Gas             & 9   & Y \\ 
           &                  &                                              &                  &$6.84^{\tiny{+0.08}}_{\tiny{-0.09}}$ & $M$--$\sigma$            & 5   &   \\ 
 NGC 0253  & $ 17.9 \pm 2.0 $ & 2MASS 1.3m (NED$^a$)                         & $K_s$            &$7.01^{\tiny{+0.30}}_{\tiny{-0.30}}$ & Maser Modeling        & 10  & Y \\ 
 NGC 0753  & $ 13.2 \pm 0.6 $ & INT 2.5m (NED$^a$)                           & $B$                &$7.22^{\tiny{+0.08}}_{\tiny{-0.09}}$ & $M$--$\sigma$            & 5   & Y \\ 
 NGC 1068  & $ 20.6 \pm 4.5 $ & UKSchmidt (NED$^a$)                          & 468 nm           &$6.95^{\tiny{+0.02}}_{\tiny{-0.02}}$ & Maser Modeling        & 11  & Y \\
 NGC 1300  & $ 10.3 \pm 1.8 $ & 2.5 m du Pont (CGS$^c$)                      & $B$                &$7.85^{\tiny{+0.29}}_{\tiny{-0.29}}$ & Stars/Gas             & 12  & Y \\ 
 NGC 1353  & $ 13.7 \pm 2.3 $ & 2MASS 1.3m (NED$^a$)                         & $K_s$            &$6.68^{\tiny{+0.08}}_{\tiny{-0.09}}$ & $M$--$\sigma$            & 5   & Y \\ 
 NGC 1357  & $ 11.8 \pm 4.8 $ & 2.5 m du Pont (CGS$^c$)                      & $B$                &$7.22^{\tiny{+0.08}}_{\tiny{-0.09}}$ & $M$--$\sigma$            & 5   & Y \\ 
 NGC 1417  & $ 12.9 \pm 4.1 $ & 2.5 m du Pont (CGS$^c$)                      & $V$                &$7.62^{\tiny{+0.08}}_{\tiny{-0.10}}$ & $M$--$\sigma$            & 5   & Y \\ 
 NGC 2273  & $ 17.5 \pm 7.2 $ & KPNO 2.1m (NED$^a$)                          & $K$                &$6.88^{\tiny{+0.02}}_{\tiny{-0.02}}$ & Maser Modeling        & 13  & Y \\ 
 NGC 2639  & $ 12.9 \pm 1.2 $ & {\it HST} (NED$^a$)                                & F606W            &$8.17^{\tiny{+0.11}}_{\tiny{-0.15}}$ & $M$--$\sigma$            & 5   & Y \\ 
 NGC 2742  & $ 32.5 \pm 7.9 $ & CAHA 2.2m (NED$^a$)                          & 1.25 $\mu$m      &$5.92^{\tiny{+0.08}}_{\tiny{-0.09}}$ & $M$--$\sigma$            & 5   & Y \\ 
 NGC 2841  & $  7.1 \pm 1.5 $ & {\it Spitzer} IRAC (NED$^a$)                       & IRAC 3.6 $\mu$m  &$8.00^{\tiny{+0.09}}_{\tiny{-0.12}}$ & $M$--$\sigma$            & 5   & Y \\ 
 NGC 2903  & $ 15.1 \pm 3.0 $ & Pal 60inch (NED$^a$)                         & 440 nm           &$6.96^{\tiny{+0.08}}_{\tiny{-0.09}}$ & $M$--$\sigma$            & 5   & Y \\ 
 NGC 2960  & $  7.5 \pm 1.7 $ & Palomar 48-inch Schmidt (NED$^a$)            & 645 nm           &$7.06^{\tiny{+0.02}}_{\tiny{-0.02}}$ & Maser Modeling        & 13  & Y \\ 
 NGC 2998  & $ 14.5 \pm 9.4 $ & KPNO 2.1m (NED$^a$)                          & 656.3 nm         &$7.08^{\tiny{+0.08}}_{\tiny{-0.09}}$ & $M$--$\sigma$            & 5   & Y \\ 
 NGC 3031  & $ 15.4 \pm 8.6 $ & {\it Spitzer} IRAC (NED$^a$)                       & IRAC 5.8 $\mu$m  &$7.91^{\tiny{+0.11}}_{\tiny{-0.07}}$ & Stars/Gas             & 14  & Y \\ 
 NGC 3145  & $  7.2 \pm 1.3 $ & 2.5 m du Pont (CGS$^c$)                      & $B$                &$7.85^{\tiny{+0.09}}_{\tiny{-0.11}}$ & $M$--$\sigma$            & 5   & Y \\ 
 NGC 3198  & $ 30.0 \pm 6.7 $ & KPNO 2.1m CFIM (NED$^a$)                     & 700  nm          &$6.10^{\tiny{+0.08}}_{\tiny{-0.09}}$ & $M$--$\sigma$            & 5   & Y \\ 
 NGC 3223  & $ 10.9 \pm 2.2 $ & 2.5 m du Pont (CGS$^c$)                      & $B$                &$7.81^{\tiny{+0.09}}_{\tiny{-0.11}}$ & $M$--$\sigma$            & 5   & Y \\ 
 NGC 3227  & $ 12.9 \pm 9.0 $ & JKT (NED$^a$)                                & H$\alpha$        &$7.33^{\tiny{+0.18}}_{\tiny{-0.10}}$ & Stars/Gas             & 15  & Y \\
           &                  &                                              &                  &$7.60^{\tiny{+0.24}}_{\tiny{-0.24}}$ & Reverberation Mapping & 16  &   \\
 NGC 3310  & $ 22.7 \pm 9.1 $ & {\it HST} WFPC2 (NED$^a$)                          & F814W            &$<7.62$                              & Stars/Gas             & 17  & Y \\ 
 NGC 3351  & $ 11.1 \pm 1.8 $ & CTIO 4.0m (NED$^a$)                          & $B$                &$<6.96$                              & Stars/Gas             & 18  & Y \\ 
 NGC 3367  & $ 36.8 \pm 5.3 $ & OAN Martir 2.12m (NED$^a$)                   & $I$                &$>5.20$                              & Eddington             & 19  & Y \\ 
 NGC 3368  & $ 14.0 \pm 1.4 $ & VATT Lennon 1.8m (NED$^a$)                   & $R$                &$6.90^{\tiny{+0.08}}_{\tiny{-0.10}}$ & Stars/Gas             &17,20& Y \\ 
 \enddata
 \end{deluxetable*}

\addtocounter{table}{-1} 
\begin{deluxetable*}{lrcccccc}
\tablecolumns{8}
\tablecaption{(Continued)}
\tablehead{
\colhead{Galaxy} & \colhead{$P$ (deg.) } & \colhead{Image Source} & \colhead{Filter} & \colhead{$\log$($M_{\rm BH}$/$M_{\odot}$)} &
\colhead{ Measurement Type} & \colhead{Source} & \colhead{Preferred} 
}
\startdata
  NGC 3393  & $ 13.1 \pm 2.5 $ & CTIO 0.9m (NED$^a$)                          & $B$                &$7.52^{\tiny{+0.03}}_{\tiny{-0.03}}$ & Maser Modeling        &21,22& Y \\ 
 NGC 3516  & $ 10.6 \pm 4.3 $ & {\it HST} WFPC2 (NED$^a$)                          & 500.7 nm         &$7.61^{\tiny{+0.18}}_{\tiny{-0.18}}$ & Reverberation Mapping & 1   & Y \\ 
 NGC 3621  & $ 12.7 \pm 1.2 $ & 2.5 m du Pont (CGS$^c$)                      & $B$                &$>3.60$                              & Eddington             & 23  & Y \\ 
 NGC 3783  & $ 10.5 \pm 4.8 $ & LCO 2.5m (NED$^a$)                           & $K$                &$7.45^{\tiny{+0.13}}_{\tiny{-0.13}}$ & Reverberation Mapping & 1   & Y \\ 
 NGC 3938  & $ 22.4 \pm 7.2 $ & KPNO 2.1m CFIM (NED$^a$)                     & $B$                &$>4.26$                              & Eddington             & 19  & Y \\ 
 NGC 3982  & $ 14.0 \pm 0.4 $ & MaunaKea2.24m (NED$^a$)                      & $R$                &$<7.93$                              & Stars/Gas             & 18  & Y \\ 
 NGC 3992  & $  6.2 \pm 6.1 $ & MaunaKea2.24m (NED$^a$)                      & $B$                &$<7.78$                              & Stars/Gas             & 18  & Y \\ 
 NGC 4041  & $ 23.3 \pm 8.2 $ & Palomar 48-inch Schmidt (NED$^a$)            & 645 nm           &$<7.33$                              & Stars/Gas             & 24  & Y \\ 
 NGC 4051  & $ 29.1 \pm 4.9 $ & MaunaKea2.24m (NED$^a$)                      & $B$                &$6.24^{\tiny{+0.12}}_{\tiny{-0.16}}$ & Reverberation Mapping & 32   & Y \\ 
 NGC 4062  & $ 12.4 \pm 1.4 $ & 1.8m Perkins (NED$^a$)                       & $B$                &$6.63^{\tiny{+0.08}}_{\tiny{-0.09}}$ & $M$--$\sigma$            & 5   & Y \\
 NGC 4151  & $ 11.8 \pm 1.8 $ & VLA (NED$^a$)                                & 21 cm            &$7.66^{\tiny{+0.05}}_{\tiny{-0.05}}$ & Stars/Gas             & 25  & Y \\
           &                  &                                              & 21 cm            &$7.64^{\tiny{+0.11}}_{\tiny{-0.11}}$ & Reverberation Mapping & 16  &   \\
 NGC 4258  & $  7.7 \pm 4.2 $ & {\it Spitzer} IRAC (NED$^a$)                       & IRAC 8.0 $\mu$m  &$7.90^{\tiny{+0.25}}_{\tiny{-0.25}}$ & Stars/Gas             & 17  & Y \\ 
           &                  &                                              &                  &$7.59^{\tiny{+0.01}}_{\tiny{-0.01}}$ & Maser Modeling        &26,27&   \\ 
           &                  &                                              &                  &$7.48^{\tiny{+0.08}}_{\tiny{-0.10}}$ & $M$--$\sigma$            & 5   &   \\
 NGC 4303  & $ 13.5 \pm 4.6 $ & 1.3m McGraw-Hill (NED$^a$)                   & $B$                &$6.92^{\tiny{+0.29}}_{\tiny{-1.14}}$ & Stars/Gas             & 17  & Y \\
  NGC 4321  & $ 21.8 \pm 3.6 $ & KP 2.1m CFIM (NED$^a$)                       & $R$                &$<7.46$                              & Stars/Gas             & 18  &   \\ 
           &                  &                                              & $R$                &$6.47^{\tiny{+0.08}}_{\tiny{-0.09}}$ & $M$--$\sigma$            & 5   & Y \\
 NGC 4388  & $ 26.2 \pm 8.2 $ & KPNO 2.3m (NED$^a$)                          & $K_s$            &$6.93^{\tiny{+0.01}}_{\tiny{-0.01}}$ & Maser Modeling        & 13  & Y \\ 
 NGC 4395  & $ 35.2 \pm 6.8 $ & {\it GALEX} (NED$^a$)                              & FUV              &$5.56^{\tiny{+0.12}}_{\tiny{-0.16}}$ & Reverberation Mapping & 28  & Y \\ 
 NGC 4450  & $  9.1 \pm 3.1 $ & KP 2.1m CFIM (NED$^a$)                       & $B$                &$<8.07$                              & Stars/Gas             & 18  & Y \\ 
 NGC 4501  & $ 12.7 \pm 1.4 $ & KP9 t2ka (NED$^a$)                           & $R$                &$<7.98$                              & Stars/Gas             & 18  & Y \\
  NGC 4536  & $ 14.8 \pm 7.9 $ & KP 2.1m CFIM (NED$^a$)                       & $B$                &$>3.68$                              & Eddington             & 19  & Y \\ 
 NGC 4548  & $ 25.6 \pm 6.6 $ & JKT (NED$^a$)                                & $B$                &$<7.55$                              & Stars/Gas             & 18  & Y  \\ 
  NGC 4593  & $ 20.2 \pm 2.7 $ & 2.5 m du Pont (CGS$^c$)                      & $I$                &$6.97^{\tiny{+0.14}}_{\tiny{-0.14}}$ & Reverberation Mapping & 33   & Y \\ 
 NGC 4800  & $ 21.5 \pm 3.2 $ & KP9 t2ka (NED$^a$)                           & $R$                &$<7.53$                              & Stars/Gas             & 18  & Y \\ 
 NGC 5033  & $ 16.5 \pm 5.6 $ & KP 2.1 CFIM (NED$^a$)                        & $B$                &$7.24^{\tiny{+0.08}}_{\tiny{-0.09}}$ & $M$--$\sigma$            & 5   & Y \\
 NGC 5055  & $ 14.9 \pm 6.9 $ & {\it Spitzer} IRAC (NED$^a$)                       & IRAC 5.8 $\mu$m  &$6.90^{\tiny{+0.08}}_{\tiny{-0.09}}$ & $M$--$\sigma$            & 5   & Y \\ 
 NGC 5495  & $ 27.8 \pm 1.2 $ & UK 48-inch Schmidt (NED$^a$)                 & 468 nm           &$7.03^{\tiny{+0.18}}_{\tiny{-0.30}}$ & Maser Modeling        & 21  & Y \\
 NGC 5548  & $ 15.0 \pm 2.5 $ & {\it HST} (NED$^a$)                                & F606W            &$7.80^{\tiny{+0.10}}_{\tiny{-0.10}}$ & Reverberation Mapping & 34,35   & Y \\
 NGC 6323  & $ 11.8 \pm 3.4 $ & Palomar 48-inch Schmidt (NED$^a$)            & 645 nm           &$6.97^{\tiny{+0.00}}_{\tiny{-0.00}}$ & Maser Modeling        & 13  & Y \\ 
 NGC 6926  & $ 17.5 \pm 5.5 $ & 2MASS (NED$^a$)                              & $K_s$            &$6.77^{\tiny{+0.26}}_{\tiny{-0.74}}$ & Maser Modeling        & 29  & Y \\ 
 NGC 7331  & $ 22.2 \pm 4.2 $ & {\it Spitzer} IRAC (NED$^a$)                       & IRAC 3.6 $\mu$m  &$7.50^{\tiny{+0.08}}_{\tiny{-0.10}}$ & $M$--$\sigma$            & 5   & Y \\ 
 NGC 7469  & $ 28.5 \pm 4.3 $ & {\it HST} NIC2 (NED$^a$)                           & F110W            &$<7.73$                              & Stars/Gas             & 16  &   \\ 
           &                  &                                              &                  &$7.06^{\tiny{+0.11}}_{\tiny{-0.11}}$ & Reverberation Mapping & 1   & Y \\ 
 NGC 7582  & $ 14.7 \pm 7.4 $ & ESO 1m Schmidt (NED$^a$)                     & $R$                &$7.74^{\tiny{+0.17}}_{\tiny{-0.18}}$ & Stars/Gas             & 30  & Y \\ 
 NGC 7606  & $ 11.3 \pm 1.2 $ & 2.5 m du Pont (CGS$^c$)                      & $V$                &$7.27^{\tiny{+0.08}}_{\tiny{-0.09}}$ & $M$--$\sigma$            & 5   & Y \\ 
 UGC 3789  & $ 10.5 \pm 4.8 $ & Palomar 48-inch Schmidt (NED$^a$)            & 645 nm           &$6.96^{\tiny{+0.30}}_{\tiny{-0.26}}$ & Maser Modeling        & 31  & Y \\
\enddata
\tablecomments{Column  1: galaxy  name.  Column  2: pitch  angle
  (degrees);  the   pitch  angle   for  the   Milky  Way   comes  from
  \citet{Levine2006}. Column 3: image source $a$ Images taken from NASA
  Extragalactic  Database   (NED).  $b$   Images  provided   by  Misty
  Bentz. $c$ The pitch angles  for Carnegie-Irvine Galaxy Survey (CGS)
  galaxies come from \citet{Davis2012}.  Column 4: image filter. Column
  5  $\log(M_{\rm  BH}$/$M_{\odot}$).  Column   6:  measurement
  type.  Column   7:  mass  measurement  source.   Column  8:  preferred
  measurement. 
\emph{{\bf References}}:
(1) \citealt{Peterson2004};
(2) \citealt{Bentz:2009};
(3) \citealt{Greenhill:2003b};
(4) \citealt{Boker1999};
(5) \citealt{Ferrarese2002};
(6) \citealt{Ishihara:2001};
(7) \citealt{Bender:2005};
(8) \citealt{Merritt2001};
(9) \citealt{Gillessen:2009};
(10) \citealt{Rodriguez-Rico:2006};
(11) \citealt{Lodato:Bertin:2003};
(12) \citealt{Atkinson:2005};
(13) \citealt{Kuo:2011};
(14) \citealt{Devereux:2003};
(15) \citealt{Davies:2006};
(16) \citealt{Hicks:Malkan:2008};
(17) \citealt{Pastorini:2007};
(18) \citealt{Sarzi:2002};
(19) \citealt{Satyapal2008};
(20) \citealt{Nowak:2010};
(21) \citealt{Kondratko:2006};
(22) \citealt{Kondratko2008};
(23) \citealt{Satyapal2007};
(24) \citealt{Marconi:2003};
(25) \citealt{Onken:2007};
(26) \citealt{Herrnstein:2005};
(27) \citealt{Miyoshi1995};
(28) \citealt{Peterson:2005};
(29) \citealt{Greenhill:2003a};
(30) \citealt{Wold:2006};
(31) \citealt{Braatz:Gugliucci:2008};
(32) \citealt{Denney2010};
(33) \citealt{Denney2006};
(34) \citealt{Bentz2007};
(35) \citealt{Bentz2009B}.}
\label{table:data}
\end{deluxetable*}
includes spiral galaxies that
have measurable spiral arm pitch  angles and measured SMBH masses. For
many of these  galaxies, we also measure the S\'ersic  index to compare
with the results of \citet{Graham2007}.

\subsection{Sample Selection} 
\label{sec:observations}

The  most desirable  sample for  us  to use  in this  analysis is  the one
consisting of  spiral galaxies  with direct  measurements of  the SMBH
mass through examination  of either stellar or gas  dynamics, or both,
within the sphere of influence of the nuclear SMBH.  We have available
$10$  galaxies  with  masses  measured  in  this  way.  There  are  an
additional  $12$ galaxies  that have  upper limits  on their  measured
masses from stellar  or gas dynamics and have no  other estimations of
their masses using other direct  techniques as discussed below.  We do
not make use of these limits in constructing our relation, but we will
later discuss the extent to which these limits are compatible with it.
Nearly all ($7$ of $10$) of these black hole mass measurements are for
relatively large black  holes, with masses $M_{\rm  BH} > 1\times10^7$
$M_{\odot}$  in galaxies  with relatively  tightly wound  spiral arms,
with typically  $P<15^{\circ}$. If we  notice that the mean  value for
pitch angle  in nearby  spiral galaxies  is $21.44^{\circ}$ \citep{Davis2013},  then  all  but  one  of  these  measurements,
excluding limits, are for galaxies  whose spiral arms have pitch angle
less than  this average.  This  means that  we are missing  the entire
right-hand side  of the distribution.  That is to  say, we have little
information on the correlation for loosely wound spirals.  It has been
argued  that spirals  with the  smallest black  holes, i.e.   the most
loosely   wound  spirals,   may  not   fit  any   correlation  without
considerable    scatter    \citep{Kormendy2011A},   but    substantial
improvement in  the relation would be  possible with more data  in the
region  from $20^{\circ}$ to $30^{\circ}$ in  pitch angle.   There  is even  more
evidence that  some of these galaxies  may not contain black  holes at
all, but  merely nuclear star  clusters (or black holes  within larger
nuclear star  clusters).  This,  however, is beyond  the scope  of our
present discussion.

Fortunately,   other  techniques   are  available   which  provide   a
considerable  number of  further data  points in  order to  expand our
sample. These  techniques provide  several galaxies with  smaller mass
black holes than are available from  the methods which can be employed
in normal  galaxies.  We  will incorporate  data available  from maser
modeling and reverberation mapping.

The   maser  modeling   data  come   from  \citet{Lodato:Bertin:2003,
  Pastorini:2007,   Ishihara:2001,    Kondratko:2006,   Kondratko2008,
  Rodriguez-Rico:2006,     Greenhill:2003a,     Braatz:Gugliucci:2008,
  Greenhill:2003b, Kuo:2011}.   Observations of  H$_2$O masers  in the
vicinity of an active black hole are used to obtain circumnuclear disk
rotation curves which  can generate accurate measurements  of the mass
of SMBHs in these galaxies.

We  also use  $14$  galaxies with  mass  estimates from  reverberation
mapping,  $12$ of  which have  not  been measured  using other  direct
methods.   For discussions  of this  method see  \citet{Peterson:2005,
  Bentz2009}.

Taking  these  three  categories together  (direct  measurements  from
stellar or gas dynamics, maser  modeling and reverberation mapping) we
have a final sample which  includes $10$ measurements using stellar or
gas  dynamics,  $12$   using  maser  modeling  data,   and  $12$  from
reverberation mapping. Where the three  samples overlap, we choose one
of  them  as  our  preferred  value  (see  Table~\ref{table:data}  for
details). This  gives us a final  sample of $34$ spiral  galaxies with
direct measurements of their central black hole masses.

As  a check  on our  work,  we will  consider  a further  data set  of
galaxies  with more  indirect  measurements in  our discussion.   This
includes $4$  galaxies with lower  limits set by the  Eddington Limit,
and $23$  ($3$ also with  direct measurements) with black  hole masses
estimated       by       the        $M$--$\sigma$       relation       of
\citet{Ferrarese2002}. Although there are  more recent publications on
the  $M$--$\sigma$ relation,  we  use  this source  for  the purposes  of
drawing a  comparison with our own  previous work and postpone  a more
thorough  discussion of  the  specific relation  between $\sigma$  and
pitch  angle to  a future  work. See  Table~\ref{table:data} for  full
details on  this extended data set.   Also, please note that  there is
an overlap between some of the techniques mentioned above.

In certain  cases we have  multiple mass estimates or  measurements of
galaxies in our data set.  Table~\ref{table:data} indicates which mass
estimates  we select  for individual  galaxies. Here  we favor  direct
measurement  techniques,  such  as   stellar  or  gas  dynamics,  over
techniques, such  as maser  modeling and reverberation  mapping, which
depend upon the black hole being active.

Whether maser modeling and reverberation mapping should be placed in a
different category from other direct  techniques is, of course, highly
debatable. Clearly, they belong to a class of techniques which observe
signals from  material in direct  orbit around the black  hole itself,
rather than  with techniques  such as  the $M$--$\sigma$  relation, which
merely correlate  the mass of  the black hole  to some feature  of the
host galaxy. But for the purposes of this paper, it is useful to place
maser modeling  and reverberation mapping  in a category  together for
two reasons.  First, because both  methods work exclusively  for AGN
and there  has been a  recent claim  that the $M$--$\sigma$  relation, at
least, is different for AGN than for normal galaxies \citep{Park2012}.
Second, because  these two methods  cover a much greater  stretch of
the sample space  than the other available direct  methods, which tend
to  have had  success  exclusively for  galaxies  containing the  most
massive black holes.

A  few exceptions  to favoring  direct stellar/gas  based measurements
exist. In  cases where  these are  available we  have chosen  the most
recent and reliable mass estimates available for the galaxy from among
a  variety of  measurement techniques.   Where these  only produce  an
upper  limit  but  not  an  estimated  mass,  we  have  chosen,  where
available, mass  determinations from  another method which  provides a
measurement of  the mass with  errors instead  of just a  limit. These
differences     are    noted     in    Table~\ref{table:data}.      In
Table~\ref{table:data},  we include  multiple entries  for each  object
which  has  multiple  different  measurement types  available  in  the
literature. The measurements we prefer are  labeled in the final  column of
the table.

Besides these general choices other exceptions are also made. NGC 5055
has   a    direct   mass   estimate   in    \citet{Gultekin2009}   and
\citet{Blais-Ouellette2004}.   \citet{Gultekin2009} suggests  that the
modeling used  in the mass  determination is very uncertain.   In this
case we have chosen to fall back on the mass of the black hole derived
from the $M$--$\sigma$ relation in \citet{Ferrarese2002}.  In the case of
NGC 4395, we have decided to use the more recent reverberation mapping
data and corresponding  mass estimate over the mass  estimate based on
combining   upper   and   lower   limits    set   on   the   mass   in
\citet{Filippenko2003}.

\subsection{Imaging Data} 
\label{sec:imdat}

The majority of  images used for measuring pitch angles  came from the
NASA/IPAC                    Extragalactic                    Database
(NED) \footnote{\url{http://ned.ipac.caltech.edu/}}.  This resulted in
the implementation of a wide range of wavelength images; anywhere from
far-ultraviolet    (FUV)    to    $21$    cm    H    emission.     See
Table~\ref{table:data} for details on  each individual galaxy. Despite
this  broad range  of  wavelength imaging,  recent  results show  that
galactic pitch angle measurements are independent of the wavelength of
the  image  \citep{Seigar2006,Davis2012},  or at  least  not  strongly
dependent  \citep{Grosbol1998}  ,  especially  in the  UV  to  near-IR (NIR)
wavelength regimes.  Although this correlation  has not been tested in
the mid-IR to  radio regimes, we can assume that  it still applies for
several reasons.   For the mid-  IR regime,  i.e., where we  have used
{\it Spitzer}/IRAC  5.8 $\mu$m and 8.0 $\mu$m imaging,  star formation
is being traced,  and so this should  give a similar pitch  angle to a
$B$-band image.  In two cases we resort to 21$cm$ data to measure pitch
angles.  While  there is  currently no empirical  evidence correlating
21 $cm$ pitch  angles to optical  or NIR  pitch angles, there  is no
reason to suggest  that the 21 $cm$ data are not  influenced in a similar
way by the underlying density wave.

When given the  option, the imaging with the best  resolution was used
(typically  $B$-band  images).   According  to  \citet{Thornley1996},  a
spiral that appears flocculant in the $B$ band may appear to have a weak
grand  design spiral  in  the NIR.   In these  cases,
NIR    imaging   was    investigated   (typically    from   2MASS,
\citealt{Jarrett2000}, or {\it Spitzer} images).

Some   galaxies  in   our  sample   had  pitch   angles  measured   in
\citet{Davis2012}, where  the method of  measuring pitch angle  we use
here is  described.  Some of  these previously measured  pitch angles,
reported  in \citet{Davis2012},  used  high-quality  imaging from  the
Carnegie-Irvine                      Galaxy                     Survey
\citep[CGS;\footnote{\url{http://cgs.obs.carnegiescience.edu/}}][]{Ho2011},
providing a  desirable set  of input  imaging for  our two-dimensional
Fast Fourier transform software, named 2DFFT.  Images highlighting the
sharpest  detail  in  spiral  arm structure  were  primarily  selected
(typically $B$-band images). See  \citet{Davis2012} for details on these
images.

Finally, for some galaxies with  AGN in our sample,
we   have  consulted   high-resolution   {\it HST}  ACS   F550M  images   of
reverberation-mapped  AGN host  galaxies,  used in  \citet{Bentz2009}.
Many of  these images did not  reveal any spiral structure  because of
the bright  nucleus or the small  FOV.  In these cases  we resorted to
ground-based, wider FOV  images to measure pitch angles.   In the end,
three  galaxies  had  pitch  angles   measured  using  {\it HST}  data  from
\citet{Bentz2009}.

\vspace{2mm}

\subsection{Measuring Pitch Angles} 
\label{sec:model}

We use the method described in \citet{Davis2012} to accurately measure
the pitch  angles of the  $34$ galaxies with direct  mass measurements
that comprise our  sample, plus additional $33$  galaxies with mass
limits or  $M$--$\sigma$ estimates used  in our extended data  set.  This
technique   is    an   extension   of   the    method   described   in
\citet{Saraiva1994}, which  utilizes a 2DFFT algorithm  to measure the
pitch  angle between  a  user  defined inner  and  outer  radius on  a
deprojected  image  of   a  spiral  galaxy.   For   more  details,  see
\citet{Puerari1992,   Puerari2000}.   Galaxies   are  deprojected   by
assuming  that  the disk  galaxy,  when  face-on, will  have  circular
isophotes.  Although \citet{Ryden2004} has shown that disk galaxies do
have   an  intrinsic   ellipticity,  it   is  relatively   small,  and
\citet{Davis2012} have  shown that small  errors in the  measured axial
ratio of galaxies do not affect the measured spiral arm pitch angle.

The extension to this method  in \citet{Davis2012} eliminates the user
defined inner  radius in favor of  measuring the pitch angle  over all
possible inner radii, thus allowing the user to examine the results of
the Fourier  analysis for long  regions over which changing  the inner
radius  of the  transformed region  of the  data does  not affect  the
measured pitch angle.   This provides a far  more accurate measurement
of the  pitch angle of  the galaxy than single  measurement techniques
utilizing individual inner  radii, as well as providing us  a means of
examining the consistency  of the logarithmic structure  of the spiral
arms.  For further details on this technique, see \citet{Davis2012}.

There are  many advantages for using  this method.  First, it  helps us
eliminate a great deal of  uncertainty involved in measuring the pitch
angles  of galaxies.   Instead of  simply assuming  that the  galaxy's
spiral is logarithmic,  it provides some check on the  extent to which
that is true by varying the  region over which the Fourier analysis is
performed.  If  a departure from  logarithmic behavior is  found (most
commonly in  a change of  the spiral arm  structure in the  very outer
regions of  the galaxy) then the  user can select a  region over which
the behavior  is consistent  and use the  pitch angle  associated with
that region. In our case this means  we are able to focus on the inner
region where we  expect that the behavior of the  spiral arm structure
will  be   more  strongly  affected   by  the  mass  of   the  central
concentration.     Additionally,     both    \citet{Seigar2006}    and
\citet{Davis2012}  established  the  consistency  between  $B$-band  and
NIR-band pitch angles, demonstrating that  pitch angle does not depend
measurably  on   the  band  chosen   for  imaging.   The   results  of
\citet{Davis2012} demonstrate that the pitch angles of the galaxies in
that sample  appear to  be generally  independent of  the band  of the
observations, at least within the  uncertainties of the reported pitch
angles.   Only  one  group  have   so  far  reported  such  an  effect
\citep{Grosbol1998} and  they agree that  the amount of  the variation
(which is visible in only three  of their studied galaxies) is no more
than seven degrees between $K$-band  and $B$-band images.  Thus, with this
method  one may  measure the  spiral arm  pitch angle  for nearly  all
late-type galaxies, limited by little more than the requirement, which
should be random and unbiased, that the galaxy is not close to edge-on
to our line of sight.

One obvious variation in the appearance  of spiral arm galaxies is the
distinction between  flocculant and  grand design  spirals. Flocculant
spirals lack  the regularity  in arm segments  seen in  other spirals.
Because of this lack of long stretches of continuous arms, measuring a
pitch angle for them is not  so straightforward, at least using manual
methods, as it  is in the case of grand  design spirals.  However, our
method permits the user to establish whether there is, nevertheless, a
consistent spiral pattern,  with measurable pitch angle  and number of
arms, over the whole  disk. Only in a handful of  cases in this sample
does  a  flocculant  galaxy  present particular  difficulties  to  our
method, as evidenced by a larger than usual error quoted.

It  has   been  argued  (for  instance   by  \citealt{Seiden1982})  that
flocculant spiral arms may be produced by a different physical process
than   grand  design   spiral   arms   (regions  of   self-propagating
star-formation acted upon by differential rotation in the former case,
spiral  density waves  in  the  latter). It  is  also noteworthy  that
galaxies which  are flocculant in the  $B$ band may show a  grand design
pattern  in the  $R$ band  \citep{Thornley1996} due  to the  potentially
different origins of  the spiral structure in  flocculant galaxies and
the  old stellar  population in  the redder  bands tracing  the spiral
density waves \citep{Seigar1998}.  It appears that though there may be
a large difference in detailed appearance (thus in one band, the image
seems flocculant, but not in another) there are only small differences
in measurable structure, the spiral arm pitch angle. This is certainly
suggestive of a  similar underlying cause for  the different varieties
of spiral  structure. In our case,  where we did find  large errors in
the  measurement  of  flocculant  spirals we  have  preferred  to  use
NIR  images  for  such   galaxies,  which  seems  to  reduce
measurement errors noticeably.

\subsection{S\'ersic Index} 
\label{sub:sersic}

S\'ersic  index  \citep{Sersic1963}  has   been  proposed  as  another
observable feature which correlates with SMBH mass \citep{Graham2007}.
Like pitch angle,  it can be measured using only  imaging data. It may
be that a combination of these two approaches, pitch angles for spiral
galaxies and  S\'ersic index for ellipticals,  lenticulars, and edge-on
spirals will  enable observers to  estimate the SMBH mass  function in
normal  galaxies out  to  considerable  distances.  Although  S\'ersic
indices  can be  measured for  face-on  spirals, doing  so involves  a
complex process of disentangling bulge from disk and bar components of
the  galaxy. It  is  likely that  measuring the  pitch  angle of  such
galaxies will be easier and more accurate.  However, any such campaign
will certainly  demand some  analysis of how  well the  two approaches
agree in their estimate of central  black hole mass.  In order to make
a comparison  between our pitch-angle-derived masses and the  work of
\citet{Graham2007},  we  must calculate  the  S\'ersic  index for  the
galaxies in  our sample.   We have  done so for  $31$ galaxies,  and
further  $4$ have  been taken  from the  literature \citep{Graham2007,
  Kent1991,  Nowak2010}.  We have  excluded  AGN  from the  sample  of
galaxies  with  measured  S\'ersic   index  due  to  the  difficulties
presented by the bright nucleus.

The S\'ersic profile relates how the  brightness of a galaxy falls off
with distance from the center. It is of the form:
\begin{equation}
I(R) = I_e\exp^{-b_n[(R/R_e)^n-1]},
\end{equation}
where $R$  is the radius  of the isophote,  $R_e$ is the  radius which
encloses half  of the light of  the galaxy, $I_e$ is  the intensity at
this radius, $b_n$ is fitted with the function
\begin{equation}
b_n = 1.9992n - 0.3271 
\end{equation}
\citep{Graham2005},  and  lastly,  $n$  is the  S\'ersic  index.   The
S\'ersic index is  also a measure of the concentration  of the galaxy,
defined  as the  amount of  light enclosed  by some  fraction (usually
taken to  be around a  third) of the  effective radius divided  by the
amount of light enclosed by  the effective radius, which by definition
is half the light of the bulge.

We  measure  the S\'ersic  index  by  fitting the  surface  brightness
contours  produced  by  Image  Reduction  and  Analysis
Facility  routine \emph{Ellipse}  \citep{Tody1986, Jedrzejewski1987}.
A bulge/disk  decomposition is performed  on these data and  a S\'ersic
profile (plus  exponential disk  for spirals) is  fit to  the resultant
data.

It was shown in \citet{Graham2001} that light concentration correlates
strongly with black  hole mass. Later, \citet{Graham2007}  found a log
relation between S\'ersic index and  black hole mass. This correlation
is of the form
\begin{equation}
\begin{split}
\log(M_{\rm bh}) = (7.98\pm0.09)+(3.70\pm0.46)\log(n/3)\\-(3.10\pm0.84)[\log(n/3)]^2.
\end{split}
\end{equation}
Thus the  S\'ersic index provides  an estimate of SMBH  masses through
images  of galactic  bulges.   We will  make  comparisons between  our
technique    and     the    results    of     \citet{Graham2007}    in
Section~\ref{sub:sersicres}.   The   S\'ersic  indices   for  several
galaxies in our  sample, along with corresponding  mass estimates from
the    relation     of    \citet{Graham2007}    are     included    in
Table~\ref{table:sersic}.
\begin{deluxetable}{ccccccc}
\tabletypesize{\scriptsize}
\tablecaption{Sample Morphology and S\'ersic Index}
\tablehead{
\colhead{Galaxy} & \colhead{ Morphology } & \colhead{S\'ersic Index} & \colhead{$\log$($M_{\rm BH}$/$M_{\odot}$) } & \colhead{Source} 
}
\startdata
    IC 2560 &  (R\arcmin)SB(r)b    & $ 1.19^{\tiny{+0.24}}_{\tiny{-0.20}}$ & $      6.994^{\tiny{+0.045}}_{\tiny{-0.077}}    $ &  1 \\ 
        M31 &  SA(s)b        & $ 3.19^{\tiny{+0.64}}_{\tiny{-0.53}}$ & $      8.081^{\tiny{+0.331}}_{\tiny{-0.327}}    $ &  1 \\  
  Milky Way &  S?            & $ 1.32^{\tiny{+0.22}}_{\tiny{-0.26}}$ & $      7.055^{\tiny{+0.081}}_{\tiny{-0.067}}    $ &  2 \\    
   NGC 0253 &  SAB(s)c       & $ 1.17^{\tiny{+0.23}}_{\tiny{-0.19}}$ & $      6.985^{\tiny{+0.039}}_{\tiny{-0.073}}    $ &  1 \\ 
   NGC 0753 &  SAB(rs)bc     & $ 1.58^{\tiny{+0.32}}_{\tiny{-0.26}}$ & $      7.190^{\tiny{+0.093}}_{\tiny{-0.119}}    $ &  1 \\ 
   NGC 1357 &  SA(s)ab       & $ 1.48^{\tiny{+0.30}}_{\tiny{-0.25}}$ & $      7.137^{\tiny{+0.083}}_{\tiny{-0.107}}    $ &  1 \\ 
   NGC 1417 &  SAB(rs)b      & $ 1.56^{\tiny{+0.31}}_{\tiny{-0.26}}$ & $      7.179^{\tiny{+0.092}}_{\tiny{-0.113}}    $ &  1 \\ 
   NGC 2903 &  SAB(rs)bc     & $ 3.28^{\tiny{+0.66}}_{\tiny{-0.55}}$ & $      8.128^{\tiny{+0.347}}_{\tiny{-0.340}}    $ &  1 \\ 
   NGC 2998 &  SAB(rs)c      & $ 2.27^{\tiny{+0.45}}_{\tiny{-0.38}}$ & $      7.577^{\tiny{+0.199}}_{\tiny{-0.203}}    $ &  1 \\ 
   NGC 3031 &  SA(s)ab       & $ 3.23^{\tiny{+0.54}}_{\tiny{-0.65}}$ & $      8.102^{\tiny{+0.401}}_{\tiny{-0.279}}    $ &  3 \\  
   NGC 3145 &  SB(rs)bc      & $ 1.58^{\tiny{+0.32}}_{\tiny{-0.26}}$ & $      7.190^{\tiny{+0.093}}_{\tiny{-0.119}}    $ &  1 \\ 
   NGC 3198 &  SB(rs)c       & $ 1.33^{\tiny{+0.27}}_{\tiny{-0.22}}$ & $      7.060^{\tiny{+0.060}}_{\tiny{-0.091}}    $ &  1 \\ 
   NGC 3227 &  SAB(s)a pec   & $ 2.52^{\tiny{+0.50}}_{\tiny{-0.42}}$ & $      7.718^{\tiny{+0.235}}_{\tiny{-0.236}}    $ &  1 \\ 
   NGC 3310 &  SAB(r)bc pec  & $ 1.89^{\tiny{+0.38}}_{\tiny{-0.32}}$ & $      7.362^{\tiny{+0.144}}_{\tiny{-0.156}}    $ &  1 \\ 	
   NGC 3351 &  SB(r)b        & $ 2.40^{\tiny{+0.48}}_{\tiny{-0.40}}$ & $      7.651^{\tiny{+0.217}}_{\tiny{-0.223}}    $ &  1 \\ 
   NGC 3367 &  SB(rs)c       & $ 0.98^{\tiny{+0.20}}_{\tiny{-0.16}}$ & $      6.914^{\tiny{+0.023}}_{\tiny{-0.065}}    $ &  1 \\ 
   NGC 3368 &  SAB(rs)ab     & $ 2.35^{\tiny{+0.00}}_{\tiny{-0.00}}$ & $      7.622^{\tiny{+0.030}}_{\tiny{-0.030}}    $ &  4 \\  
   NGC 3621 &  SA(s)d        & $ 1.89^{\tiny{+0.38}}_{\tiny{-0.32}}$ & $      7.362^{\tiny{+0.144}}_{\tiny{-0.156}}    $ &  1 \\ 
   NGC 3938 &  SA(s)c        & $ 1.45^{\tiny{+0.29}}_{\tiny{-0.24}}$ & $      7.121^{\tiny{+0.076}}_{\tiny{-0.102}}    $ &  1 \\ 
   NGC 3982 &  SAB(r)b:      & $ 2.14^{\tiny{+0.43}}_{\tiny{-0.36}}$ & $      7.504^{\tiny{+0.180}}_{\tiny{-0.189}}    $ &  1 \\ 
   NGC 3992 &  SB(rs)bc      & $ 1.40^{\tiny{+0.28}}_{\tiny{-0.23}}$ & $      7.095^{\tiny{+0.068}}_{\tiny{-0.097}}    $ &  1 \\ 
   NGC 4041 &  SA(rs)bc      & $ 0.74^{\tiny{+0.15}}_{\tiny{-0.12}}$ & $      6.876^{\tiny{+0.015}}_{\tiny{-0.066}}    $ &  1 \\ 
   NGC 4051 &  SAB(rs)bc     & $ 2.18^{\tiny{+0.44}}_{\tiny{-0.36}}$ & $      7.527^{\tiny{+0.182}}_{\tiny{-0.196}}    $ &  1 \\ 
   NGC 4258 &  SAB(s)bc      & $ 2.04^{\tiny{+0.34}}_{\tiny{-0.41}}$ & $      7.447^{\tiny{+0.206}}_{\tiny{-0.138}}    $ &  3 \\  
   NGC 4303 &  SAB(rs)bc     & $ 0.79^{\tiny{+0.16}}_{\tiny{-0.13}}$ & $      6.877^{\tiny{+0.015}}_{\tiny{-0.064}}    $ &  1 \\ 
   NGC 4321 &  SAB(s)bc      & $ 1.86^{\tiny{+0.37}}_{\tiny{-0.31}}$ & $      7.345^{\tiny{+0.136}}_{\tiny{-0.150}}    $ &  1 \\ 
   NGC 4450 &  SA(s)ab       & $ 1.34^{\tiny{+0.27}}_{\tiny{-0.22}}$ & $      7.065^{\tiny{+0.061}}_{\tiny{-0.091}}    $ &  1 \\ 
   NGC 4501 &  SA(rs)b       & $ 2.28^{\tiny{+0.46}}_{\tiny{-0.38}}$ & $      7.583^{\tiny{+0.199}}_{\tiny{-0.209}}    $ &  1 \\ 
   NGC 4536 &  SAB(rs)bc     & $ 2.27^{\tiny{+0.17}}_{\tiny{-0.14}}$ & $      7.577^{\tiny{+0.052}}_{\tiny{-0.054}}    $ &  1 \\ 
   NGC 4548 &  SB(rs)b       & $ 1.58^{\tiny{+0.32}}_{\tiny{-0.26}}$ & $      7.190^{\tiny{+0.093}}_{\tiny{-0.119}}    $ &  1 \\ 
   NGC 4593 &  (R)SB(rs)b    & $ 2.40^{\tiny{+0.48}}_{\tiny{-0.40}}$ & $      7.651^{\tiny{+0.217}}_{\tiny{-0.223}}    $ &  1 \\ 
   NGC 5033 &  SA(s)c        & $ 1.93^{\tiny{+0.39}}_{\tiny{-0.32}}$ & $      7.385^{\tiny{+0.146}}_{\tiny{-0.163}}    $ &  1 \\ 
   NGC 5495 &  (R\arcmin)SAB(r)c   & $ 0.69^{\tiny{+0.14}}_{\tiny{-0.12}}$ & $      6.881^{\tiny{+0.016}}_{\tiny{-0.071}}    $ &  1 \\ 
   NGC 6926 &  SB(s)bc pec   & $ 1.77^{\tiny{+0.35}}_{\tiny{-0.30}}$ & $      7.295^{\tiny{+0.126}}_{\tiny{-0.138}}    $ &  1 \\ 
   UGC 3789 &  (R)SA(r)ab    & $ 0.95^{\tiny{+0.19}}_{\tiny{-0.15}}$ & $      6.905^{\tiny{+0.019}}_{\tiny{-0.063}}    $ &  1 \\ 
\enddata
\tablecomments{Column 1: galaxy name. Column 2: galaxy morphology taken from NED. Column 3: S\'ersic index. Column 4: $\log(M_{\rm BH}$/$M_{\odot}$) derived from each galaxy's measured S\'ersic index using the relation of \citet{Graham2007}. Column 5:  source of S\'ersic index measurement. {\emph {\bf References:}} (1) This work; (2) \citealt{Kent1991}; (3) \citealt{Graham2007}; (4) \citealt{Nowak2010}.}
\label{table:sersic}
\end{deluxetable}

\section{Results} 
\label{sec:results}

\subsection{An Updated SMBH Mass--Pitch Angle Relation} 
\label{sec:M-P}

As discussed above, we  chose our
sample to include spiral galaxies whose central black hole masses have
been measured using  a direct technique. We define  a direct technique
to be  one which  measures the  motions and  positions of  material in
orbit  around  the  black  hole  or  directly  within  its  sphere  of
influence.  This definition encompasses quite a few different methods.
Not all  are equally accurate or  reliable, but for our  purposes they
all have  the important distinction  that they do  extract information
from signals emitted by material in the direct gravitational influence
of the black hole.

\begin{figure}[t!]
\epsscale{1.0}
\plotone{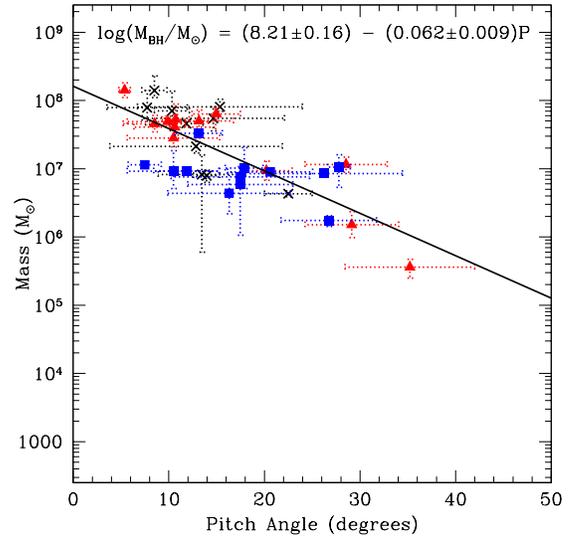}
\caption{Black hole  mass vs. pitch  angle for all  spiral galaxies
  with directly measured black hole masses available.  The best linear
  fit to  this data  is illustrated and  gives the  following relation
  $\log(M_{\rm BH}/M_{\odot}) = (8.21\pm0.16) - (0.062\pm0.009)P$.  The
  fit has  a reduced  $\chi^2 =  4.68$ with a  scatter of  $0.38$ dex.
  Black hole masses measured using stellar and gas dynamics techniques
  are labeled  with black $\times$'s (10  points), reverberation mapping
  masses with  red triangles (12 points), and  maser measurements with
  blue squares (12 points). }
\label{fig:DMR}
\end{figure}

Combining  these  three  samples,  stellar  and  gas  dynamics,  maser
modeling, and  reverberation mapping, we  have a total sample  of $34$
objects.  Fitting these data points (Figure~\ref{fig:DMR}),
%
%
%
%
we find an
updated SMBH $M$--$P$ relation of
\begin{equation}
\log(M_{\rm BH}/M_{\odot})   =   (8.21\pm0.16)-(0.062\pm0.009)P,
\end{equation}
where $P$  is the absolute  value of the  measured pitch angle  of the
galaxy  in degrees.   Please  note  that the  sign  of  a pitch  angle
measurement merely represents  the chirality of the  galaxy based upon
the  user's line of sight  and is  unimportant. This  fit has  a
$\chi^2=4.68$ and a scatter of  $0.38$ dex.  This result is consistent
with the  previous result of  \citet{SKKL08}.  It is  encouraging that,
with  significantly  more  data  available, the  scatter  has  remained
unchanged.  This  value is  less than the  scatter of  spiral galaxies
about   the   $M$--$\sigma$   relation    of   $\sim   0.56$   dex   from
\citet{Gultekin2009}.  A Pearson rank correlation coefficient test
produces a coefficient of $-0.81$, a strong anti-correlation.  This
result has a significance of $99.7\%$, a $3\sigma$ result. 

As  discussed above,  the scatter  in this  relation is  comparable to
other relationships.   A possible  reason for  the reduced  scatter in
this relationship,  when compared  with \citet{Gultekin2009},  is that
the  measurement of  $\sigma$  in spiral  galaxies  requires that  one
distinguishes the contributions of the galactic bulge from other stellar
components such  as the disk or  bar.  Where the galaxy  has an active
nucleus,  the  region  over  which $\sigma$  is  measured  is  usually
obscured and a proxy (some spectral line or lines in the AGN spectrum)
must be used. Many  of the galaxies in this sample  have AGNs in their
nuclei.

In the case  of bulge luminosity, the presence of  an AGN can sometimes
be overcome, but  when the galaxy is a disk  galaxy one must undertake
decomposition  of   the  light  from  the   galaxy,  fitting  multiple
components to the luminosity profile (bulge, disk, and bar are the main
components).   Once again,  the  presence of  unresolved nuclear  flux
(e.g.,  AGN  or  LINER  or  even nuclear  star  cluster)  will  create
difficulties in determining bulge  luminosity or S\'ersic index.  Some
algorithms have been  used to measure large numbers  of galactic bulge
luminosities for the purposes of  bulge mass estimates or measurements
of  S\'ersic index   \citep[e.g.][]{BUDDA}.  Automated  codes may  have
difficulties  disentangling  the  unresolved nucleus  and  the  bulge,
especially with ground-based seeing of  $\sim1\arcsec$ for any galaxies that
are not nearby.  The method of  \citet{Davis2012} is unaffected by this
issue.  In  the case  of measurements  of spiral  arm pitch  angle, we
examine a  component which is  unambiguous in spiral galaxies  in much
the same way that $\sigma$  is unambiguous in elliptical galaxies. The
method of \citet{Davis2012} requires minimal image processing to yield
a measurement  of the pitch  angle, requiring only a  deprojection and
cropping of  the image  and the  locations of  the central  pixel and
galactic edge.

For the purposes of estimating black hole masses on the basis of pitch
angle measurements of galactic spiral arms, we propose to use this fit
based on direct measurements of SMBH mass.

\subsection{Comparison with the Previous $M$--$P$ Relation} 
\label{sec:oldM-P}

In our  previous work \citep{SKKL08},  black hole masses  derived from
measurements of  $\sigma$, the velocity  dispersion in the  core bulge
region, all  reported in  \citet{Ferrarese2002}, were used.   This was
largely because  of the  relative scarcity  of direct  measurements of
black hole masses in spiral galaxies.  Since then the available number
of direct  measurements has  increased by  over a  factor of  two.  We
therefore exclude any $\sigma$-derived masses from the correlation fit
developed here, as it is not  a direct measurement of black hole mass.
Indeed, the correlation between $\sigma$  and black hole mass which we
use is based  upon a fit using  some of the objects we  include in our
sample.   Later   in  this   section  we   will  include   those  same
$\sigma$-derived values  in a fit  which we derive primarily  to check
how  closely our  results  agree  with the  results  reported in  that
earlier  paper  \citep{SKKL08}.   Additionally, when  we  compare  our
results with these  indirect values it provides a useful  check on the
overall validity of  our correlation. This is  especially true because
the number  of black hole  masses measured by direct  techniques which
are near or below a million solar masses is very small. Including some
$\sigma$ values does provide a useful check on the slope of the fit by
providing some  extra evidence at the  low mass end of  the graph.  In
any case the value of  our fit, including the $\sigma$-derived values,
agrees very closely with the one reported in the earlier paper.

It  is important  to note  that while  the updated  fit reported  here
differs from the one in \citet{SKKL08}, though only to a minor degree,
the discrepancy is not caused by the remeasurement of the pitch angles
by our  improved technique,  or by  revisions of  the black  hole mass
measurements used previously.  Both of  these changes were quite minor
in any case. More importantly, it is  not a new trend indicated by the
new  black  hole  mass  measurements which  were  unavailable  before.
Rather,  the difference  is purely  because we  are in  a position  to
dispose of the use of masses derived by indirect methods. It should be
understood, and it will be shown below, that the results of this paper
agree remarkably closely  with those previous results  when we include
those indirect masses  used previously.  It is only  by removing these
that we differ at all from the earlier result.

As we will argue in  Section~\ref{sec:Discussion}, we see the relation
between pitch angle and black hole mass as a natural result of density
wave theory, which  demands that the wavelength of  the spiral density
waves  should depend  directly on  the  mass of  the galaxies'  central
bulge, but  this is also true  for rival theories.  Thus  pitch angle,
S\'ersic index, $\sigma$,  and bulge luminosity all  tend to correlate
with each other because they  all indirectly measure the central bulge
mass. There is strong evidence that this in turn correlates to central
black hole mass. It may also ultimately depend on the dark matter halo
concentration in some way still to be properly elucidated.

\subsection{Subsamples of the Full Direct Sample} 
\label{sec:samp}

It is  instructive to examine  the trends in  the overall fits  of the
three different  measurement techniques we  have utilized in  our data
sample.   Below,  we consider  these  three  techniques separately  to
verify  that each  individual subsample  provides consistent  results.
Additionally,  this  approach  allows  us  to  examine  any  potential
differences  between  active  and  normal  galaxies,  as  two  of  our
subsample groups consist entirely of active galaxies.

\begin{figure}[t!]
\plotone{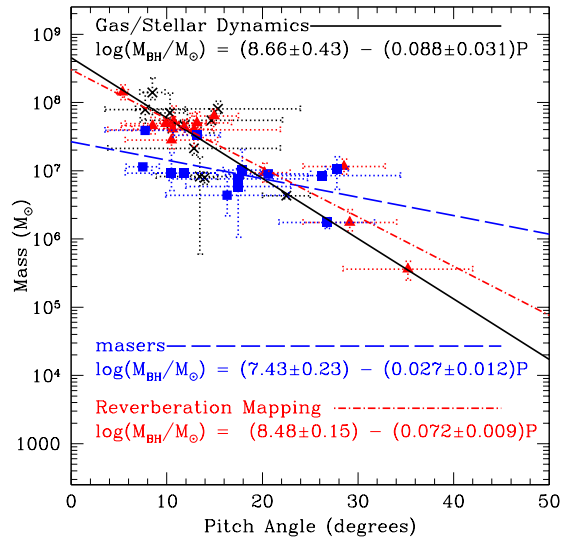}
\caption{Black hole  mass vs. pitch  angle for all  spiral galaxies
  with  directly measured  black  hole masses  available.  Black  hole
  masses  measured  using  stellar  and gas  dynamics  techniques  (10
  points)  are  labeled with  black  $\times$'s, reverberation  mapping
  masses with  red triangles (14 points), and  maser measurements with
  blue squares (13  points). The figure shows three  separate fits for
  the  three  subsamples.  The  black  solid  line is  a  fit  to  the
  gas/stellar dynamics data (black $\times$'s), the dashed blue line is
  a fit to  the maser modeling data (blue  squares), and the dot-dashed
  red  line  is   a  fit  to  the  reverberation   mapping  data  (red
  triangles).  The  gas/stellar  dynamics  fit and  the  reverberation
  mapping  fit are  fairly compatible  and both  close to  the overall
  fit.  The  maser  modeling   data,  however,  follows  a  noticeably
  shallower  slope. Given  their small  sizes  it is  difficult to  be
  certain if there is really a conflict between the subsamples.  }
\label{fig:all3}
\end{figure}

Using only the $10$ galaxies  with mass estimates utilizing stellar or
gas dynamics,  which are mostly in  normal galaxies, we find  a linear
fit to the data of the form
\begin{equation}
\log(M_{\rm BH}/M_{\odot})   =    (8.66\pm0.43)  - (0.088\pm0.031)P.
\end{equation}
This provides a $\chi^2  = 1.2$ and a scatter of  $0.39$ dex about the
linear fit  (Figure~\ref{fig:all3}).
%
%
%
%
This fit differs  from the value
of \citet{SKKL08}  when only using  mass measurements from  stellar or
gas dynamics, but  produces a consistent result, though  one with much
higher error, and is similarly consistent with our earlier results.

Examining the masers by  themselves (Figure~\ref{fig:all3}) results in
a linear fit of the form
\begin{equation}
\log(M_{\rm BH}/M_{\odot})  = (7.43\pm0.23) - (0.027\pm0.012)P.
\end{equation}
The scatter about the fit line for the masers is $0.30$ dex. These data
is comprised  of $13$ galaxies  with detectable masers, $12$  of which
have not been measured using stellar and gas dynamics methods.

As  we can  see from  Figure~\ref{fig:all3},  the maser  sample has  a
significantly shallower slope when compared  to the direct stellar and
gas dynamics measurements. This, combined with the good quality of the
fit  and low  overall  scatter, could  imply that  a  separate fit  is
necessary for this population of  active galaxies, but it is difficult
to say anything for certain with the relatively small sample size.

Turning   to   the   reverberation   mapping   subsample,   (also   in
Figure~\ref{fig:all3}) we find a linear fit of the form
\begin{equation}
\log(M_{\rm BH}/M_{\odot})  =  (8.48\pm0.15)-(0.072\pm0.009)P,
\end{equation}
 with a $\chi^2=0.93$ and a scatter of $0.28$ dex for the subsample of
 $12$ reverberation mapped masses alone.   This is consistent with the
 results of \citet{SKKL08}.  For an  AGN only sample of maser modeling
 data and reverberation mapping, we find
\begin{equation}
\log(M_{\rm BH}/M_{\odot})  =  (8.09\pm0.45)-(0.058\pm0.025)P,
\end{equation}
with a $\chi^2=3.68$ and a scatter of $0.38$ dex for this subsample.

The stellar  and gas dynamics sample  has a somewhat steeper  fit than
the  overall fit,  while  the  maser modeling  sample  has a  somewhat
shallower fit.  Interestingly, the reverberation mapping sample splits
the  difference and  ends up  very close  to the  overall slope.   The
steeper fit of the direct, normal subsample may be simply attributable
to the fact  that this sample lacks  smaller black holes and  so it is
more difficult to determine, on the  basis of this sample alone, where
the slope really lies.

The reverberation  mapping only result  shows great similarity  to the
original  fit by  \citet{SKKL08} and  is  consistent with  the fit  to
stellar and  gas dynamics data,  with much better scatter.   These are
significantly  different  fits  from   that  found  using  maser  data
above. They are also consistent with  our total sample and the results
based  on  $M_{\rm   BH}-\sigma$  measurements  as  we   will  see  in
Section~\ref{sec:M-P-M-Sig}.   Meanwhile,  the   combined  maser   and
reverberation  mapping sample  produces a  result consistent  with our
other fits,  belying any notion  that the maser results  are different
because these objects are AGN.

\subsection{Comparing $M$--$P$ Results with $M$--$\sigma$} 
\label{sec:M-P-M-Sig}

Here, we  consider one  of the most  common techniques  for estimating
galaxy black hole masses, the  $M$--$\sigma$ relation.  This technique is
used to estimate the mass of  a central SMBH by measuring the velocity
dispersion  of stars  in the  galactic bulge.   We consider  a set  of
galaxies  using the  $M$--$\sigma$  relation from  \citet{Ferrarese2002}.
This data  set utilizes a  single fit  to the $M$--$\sigma$  relation and
allows us to  fill in portions of  the righthand side  of the $M_{\rm
  BH}$-pitch angle relation.  We include $23$ galaxies  from this set,
$20$ of which do not have direct measurements. These $20$ galaxies also includes
$3$ of the galaxies for which we  have mass limits based on stellar or
gas dynamics, allowing us to replace those limits (for the purposes of
this  section) with  the mass  estimate as  derived from  the galaxy's
$\sigma$.

\begin{figure*}[t!]
\epsscale{1.0}
\plottwo{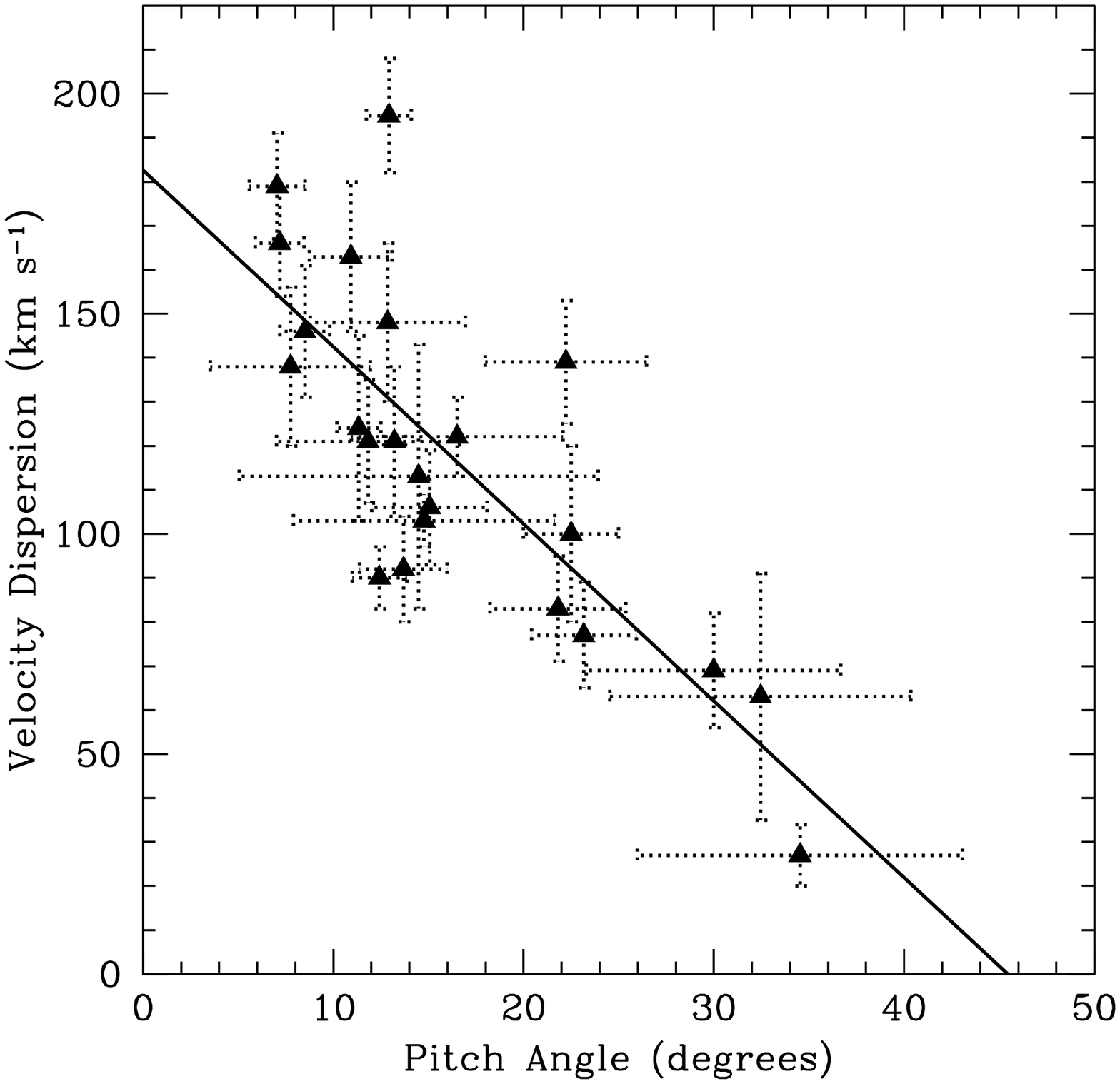}{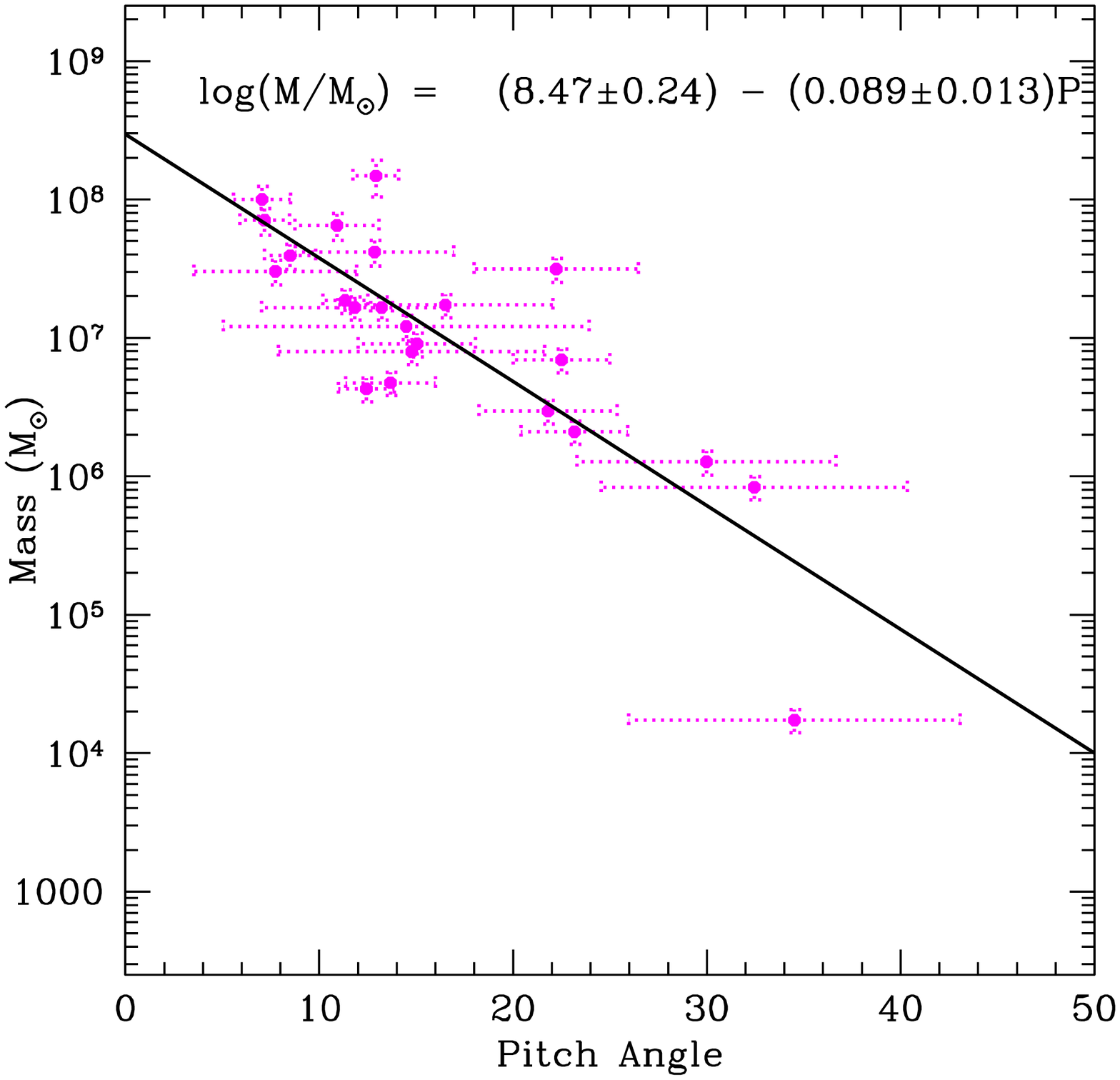}
\caption{Comparisons  of  pitch  angle  data  with  measurements of
  core stellar velocity dispersion ($\sigma$) for galaxies from
  \citet{Ferrarese2002}.   \underline{Left}: $\sigma_c$  compared with  our
  measured  pitch   angles.   \underline{Right}:  masses  taken   from  the
  $M$--$\sigma$ relation compared to spiral arm pitch angle. Both figures
  illustrate a  strong correlation  between ($\sigma$) and  spiral arm
  pitch  angle, as one would expect from our argument that both measure
  the mass of the galaxy's central bulge (see the Appendix).
  The  fit to  the  SMBH mass--pitch  angle relation  is
  $\log(M_{\rm BH}/M_{\odot}) =  (8.47\pm0.24)-(0.089\pm0.013)P$ with a
  $\chi^2=4.86$ and a scatter of $0.48$ dex.  }
\label{fig:sigma}
\end{figure*}

Figure~\ref{fig:sigma}
%
%
%
%
illustrates the  results of  comparing central
velocity dispersion ($\sigma_c$) data with spiral arm pitch angle. The
left panel of Figure~\ref{fig:sigma}  demonstrates a tight correlation
between  $\sigma_c$  and  pitch  angle.   The  right  panel  similarly
illustrates a  correlation between  the $23$  masses derived  from the
$M$--$\sigma$ relation  of \citet{Ferrarese2002}  with pitch  angle.  The
fit to the mass versus pitch angle data is
\begin{equation}
\log(M_{\rm BH}/M_{\odot})    = (8.47\pm0.24)-(0.089\pm0.013)P,
\end{equation}
with a  $\chi^2=4.86$ and  a scatter  of $0.48$  dex.  This  result is
consistent with  the earlier results  based on direct  measurements of
stellar and gas  dynamics as well as reverberation  mapping data. Note
that while the scatter is high, it is consistent with the scatter found
in \citet{Gultekin2009}  for the  total sample,  $\sim 0.44$  dex, and
lower than the result for late-type galaxies alone, $0.56$ dex.

\begin{figure}[t!]
\plotone{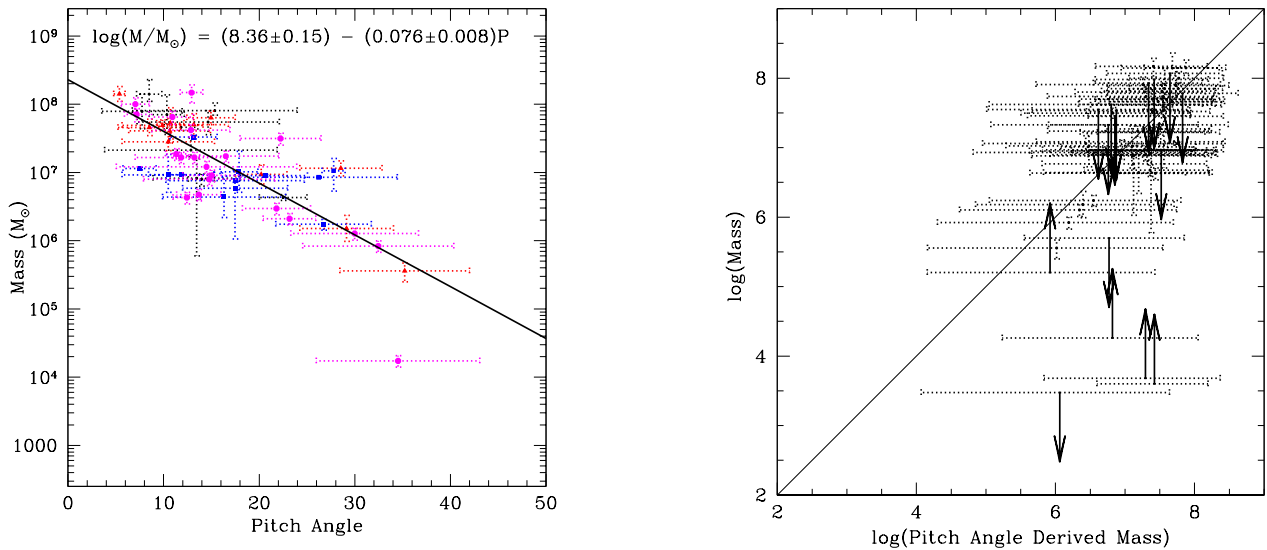}
\caption{SMBH  mass--pitch angle  relation for  all available  directly
  measured  black hole  masses  (as in  Figure~\ref{fig:DMR}) and  for
  those  masses estimated  indirectly  via $\sigma$  in our  preferred
  sample (see Table~\ref{table:data} and Section~\ref{sec:observations}
  for details).   The fit  to the SMBH  mass--pitch angle  relation for
  this    extended   data set    is   $\log(M_{\rm    BH}/M_{\odot})   =
  (8.36\pm0.15)-(0.076\pm0.008)P$ with a  $\chi^2=10.43$ and a scatter
  of $0.45$  dex. This fit is  consistent with that obtained  from our
  sample of directly measured black hole masses (Figure~\ref{fig:DMR})
  and  is almost  identical to  the fit  given in  \citet{SKKL08}. The
  addition of $\sigma$-derived masses, several of which are at the low-mass end  of our distribution, tends  to confirm the validity  of the
  complete sample  fit shown in Figure~\ref{fig:DMR}  against the much
  shallower  fit  found  for  the maser-modeling-only  data  shown  in
  Figure~\ref{fig:all3}.  Black $\times$'s represent data  from stellar
  or gas dynamics (10 points),  blue squares represent data from maser
  modeling (12 points), red  triangles come from reverberation mapping
  data (12 points) and magenta  octagons represent masses derived from
  the $M$--$\sigma$ relation (20 points).}
\label{fig:All}
\end{figure}

Taking the $\sigma$-derived masses of the $20$ galaxies without direct
measurements and  adding them to  our direct  sample of $34$  gives us
$54$  galaxies  with  mass  measurements, either  direct  or  indirect
(Figure~\ref{fig:All}).
%
%
%
%
A fit to these data points gives:
\begin{equation}
\label{eqn:final}
\log(M_{\rm BH}/M_{\odot})    =    (8.36\pm0.15)-(0.076\pm0.008)P
\end{equation}
with    a   $\chi^2=10.43$    and    a   scatter    of   $0.45$    dex
(Figure~\ref{fig:All}.)  This result is strikingly close to the result
of \citet{SKKL08}
\begin{equation}
\label{eqn:SKKL}
\log(M_{\rm BH}/M_{\odot})    =    (8.44\pm0.10)-(0.076\pm0.005)P
\end{equation}

To reiterate,  the modest  change in the  $M$--$P$ correlation  reported in
this  paper is  entirely due  to our  decision to  drop indirect  mass
measurements from our  sample, because the number  of available direct
measurements has doubled from the  previous work. Thus our results are
consistent with the earlier one  reported. This, coupled with the fact
that  the scatter  in the  relation  has improved,  suggests that  the
correlation is not simply the result of an initially small data set.

Therefore, we  have two fits  to the data  to which we  attach special
importance.  The first  is our preferred fit that  includes all direct
measurements  and which  we adopt  as  the SMBH  $M$--$P$ relation.   The
second includes  all direct data  plus values  of the black  hole mass
derived  from $\sigma_c$.   It is  heartening  that the  two fits  are
consistent  with  each other,  though  the  combined sample  has  much
tighter  constraints on  the relation  even with  its modestly  larger
scatter of $0.45$, as opposed to $0.38$ dex.

The  fit using  only direct  measurement data  has a  shallower slope,
$0.062\pm0.009$,  than  the  all-data  fit,  but  the  two  are  fully
consistent.  Although  we place greater  faith in the direct  data, we
must also acknowledge that it is  missing a significant amount of data
on the right hand side of the  relation.  Be that as it may, we prefer
to rely on the fit based only on direct measurement data for the final
result of our correlation.

\begin{figure}[t!]
\epsscale{1.0}
\plotone{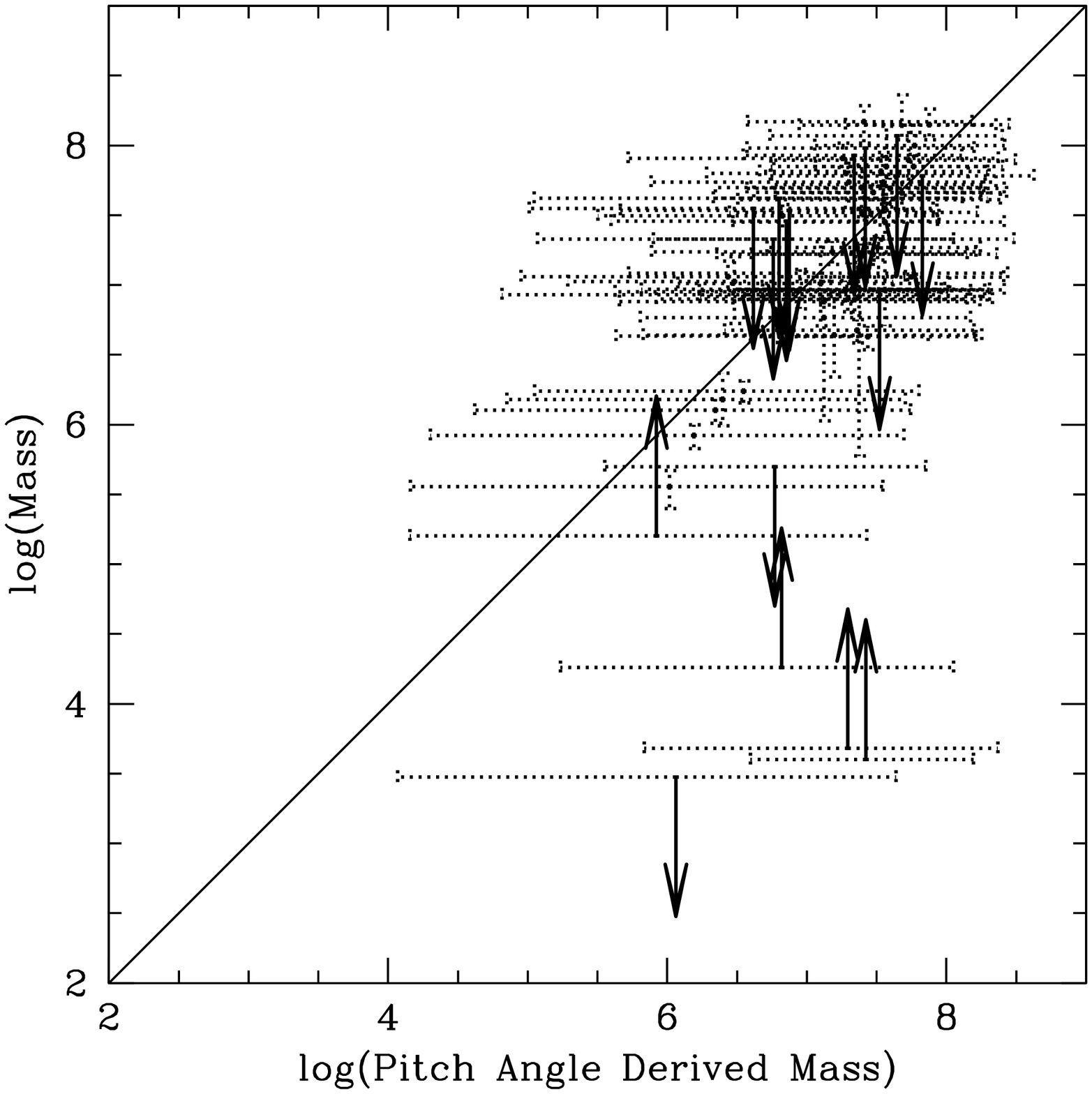}
\caption{ Measured black hole mass  of our sample of galaxies compared
  to   the  mass   derived  by   applying  the   fit   illustrated  in
  Figure~\ref{fig:DMR}  to  our measured  pitch  angles  for the  same
  objects.  The  solid black  line is included  to illustrate  the 1-1
  relation, and  the distance  of each point  from the line  gives the
  residual. It will be noticed  that some of the limits contradict our
  fit, as discussed in the text.  Two in particular, M33 and IC 342, do
  so strikingly.  }
\label{fig:PMVM}
\end{figure}

We  now examine  how consistent  the masses  generated from  the $M$--$P$
relation   are   with   the   masses   taken   from   the   literature
(Figure~\ref{fig:PMVM}).
%
%
%
%
Those  galaxies with upper and  lower limits
are included in  the right panel of  the figure. Notice that  M33 is a
distinct outlier from the rest of the data.

\subsection{Comparisons with Mass Limits} 
\label{sec:ed}

Finally we look at those galaxies  for which only limits are available
on their black hole masses.  We take  a look at the $12$ galaxies with
upper mass limits set by stellar  and gas dynamics effects, as well as
4 galaxies  which have lower limits  placed on their masses  due to
estimates  of their  luminosity in  ratio to  the Eddington  limit, to
check for  any inconsistencies of  the $M$--$P$ relation with  these mass
limits.

As  can be  seen  in  the right  panel  of Figure~\ref{fig:PMVM},  the
limits, with few exceptions, are consistent with the resulting fit. In
the cases where  the limits are not consistent, we  find that they are
either  still  consistent with  the  scatter  observed in  the  fitted
points or that  the pitch angle is at least  more consistent with the
measured $\sigma$ of the galaxy.

M33 has  direct estimates which  place an upper  limit on its  mass of
$3.0\times10^3$  $M_{\odot}$ \citep{Merritt2001}.   But note  that its
$\sigma$ suggests a  much greater mass than this, one  that is more in
line  with our  correlation (for  illustrative purposes  this $\sigma$-derived   value   of   M33's   black  hole   mass   is   included   in
Figure~\ref{fig:All}  above).  Certainly,  it  would  be reasonable  to
expect, if  there is  a discrepancy between  $\sigma$ and  more direct
black  hole  mass measurements,  that  the  pitch angle  would  follow
$\sigma$.  Similarly  IC 342  has an  upper limit  mass, based  on gas
dynamics, of $M_{\rm  BH}=5.0\times10^5$ $M_{\odot}$ which contradicts
what one  would expect from  its measured $\sigma$ (again  we included
the $\sigma$-derived value for IC 342 in Figure~\ref{fig:All} above).

The lower  limits placed on the  sample using the Eddington  limit are
also illustrated  in the  right panel of  Figure~\ref{fig:PMVM}.  This
plot does  illustrate that the  lower limits  placed on the  masses by
their Eddington  luminosities are  low, but  consistent with  the SMBH
$M$--$P$ relation.

\subsection{Accuracy of $M$--$P$ Relation for Estimating Black Hole Masses} 
\label{sub:accuracy}

In \citet{Davis2012}  we show that  pitch angle  in many cases  can be
measured  to within  an  accuracy of  $3^{\circ}$  or  lower. This  is
especially true for grand design spirals, some of which can have pitch
angle errors  of little over $1^{\circ}$. If we accept  $3^{\circ}$ as
typical of  relatively high quality  data, then the resulting  error in
black  hole mass  estimation obviously  depends  on the  slope of  the
correlation. A steep  slope will translate a modest  pitch angle error
into quite a large black hole mass error.  Fortunately, the correlation
is not especially  steep. The steepest possible  slope consistent with
the  error  in   our  preferred  correlation  is  $0.062   +  0.009  =
0.071$. Thus,  an error in pitch  angle of $3^{\circ}$  translates to a
maximum error in the  log of the black hole mass of  $0.071 \times 3 =
0.213  M_{\rm BH}/M_{\odot}$.   Thus, if  we had  a pitch  angle which
translated into  a black  hole mass of  a $1\times  10^6$ $M_{\odot}$,
such an error would mean that the mass of the black hole, as estimated
by our correlation, could vary from $1.2\times10^6$ $M_{\odot}$ on the
high end to $7.87\times10^5$ $M_{\odot}$ on  the low end or roughly $1
\pm 0.2  \times 10^6$ $M_{\odot}$  for the  case dealt with  here; the
relative error would be lower for a larger black hole.

In  the  case of  poor  quality  data  or some  especially  flocculant
galaxies, errors  can be  $10^{\circ}$  or higher.   In this  case the
error in black  hole mass is $0.71  M_{\rm BH}$ and so  the same black
hole  would have  a mass,  with  error of  $(1 \pm  0.71) \times  10^6
M_\odot$.  Therefore pitch angle measurements with errors greater than
$10^{\circ}$ will  do little better than estimating  black hole masses
to within an order of magnitude.

\subsection{S\'ersic Index} 
\label{sub:sersicres}

It was shown in \citet{Graham2001} that S\'ersic index, which measures
the light concentration of a  galactic bulge, correlates strongly with
black hole mass and in \citet{Graham2007} a quadratic relation between
the log  of the S\'ersic  index and  black hole mass  was established.
Thus,  the S\'ersic  index provides  another observational  estimate of
SMBH masses through  images of galactic bulges, and  a good competitor
to  the use  of  pitch  angle measurements  in  estimating SMBH  mass.
Certainly,  S\'ersic index  is capable  of estimating  SMBH masses  in
early-type  galaxies  where pitch  angle  is  unusable.  The  S\'ersic
indices for several  galaxies in our sample,  along with corresponding
mass estimates from the relation of \citet{Graham2007} are included in
Table~\ref{table:sersic}.   We will  examine the  quality of  S\'ersic-index-based  mass  estimates  using  these   values  as  well  as  the
morphologies of the galaxies.

\begin{figure*}[t!]
\epsscale{1.0}
\plottwo{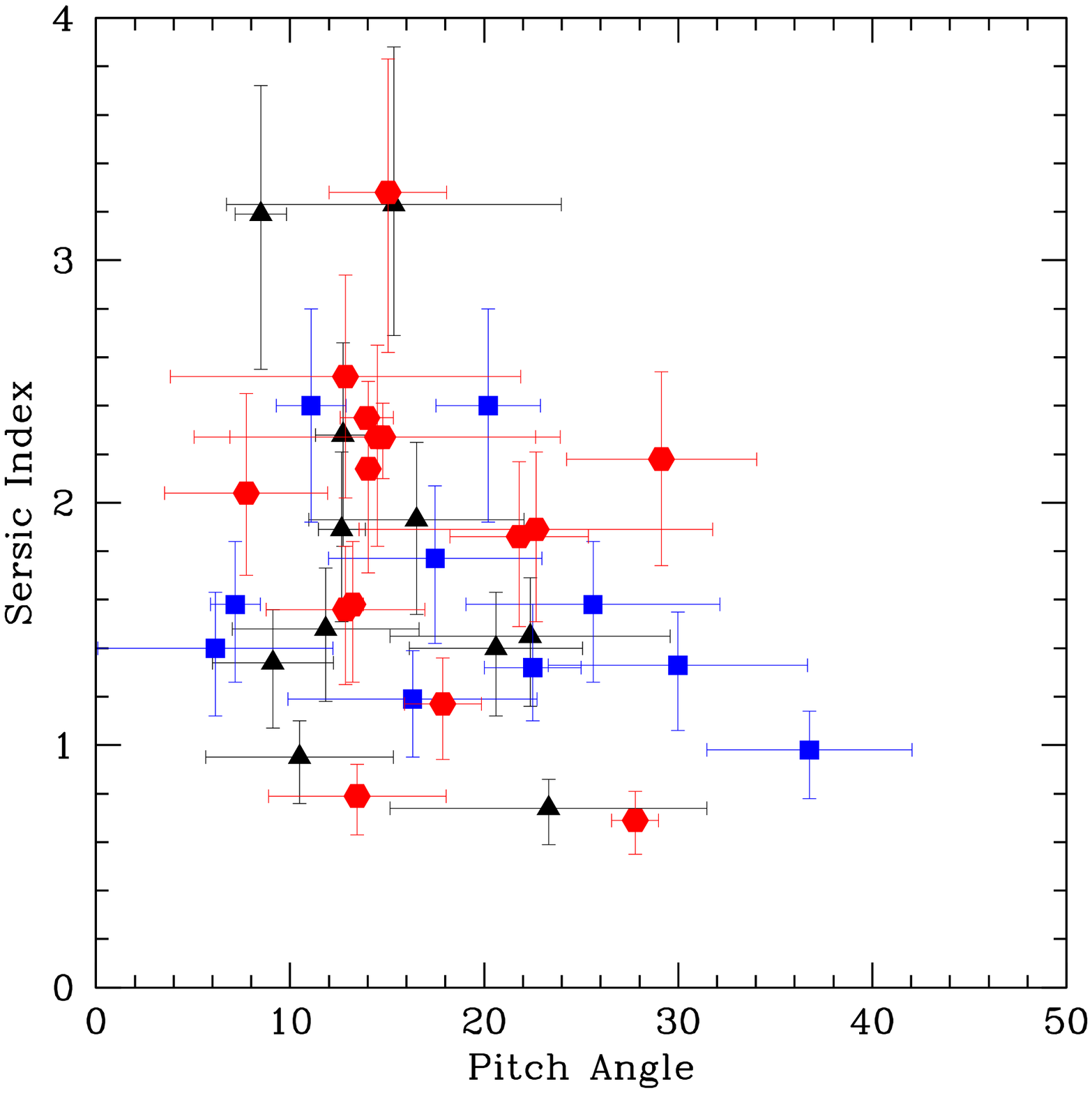}{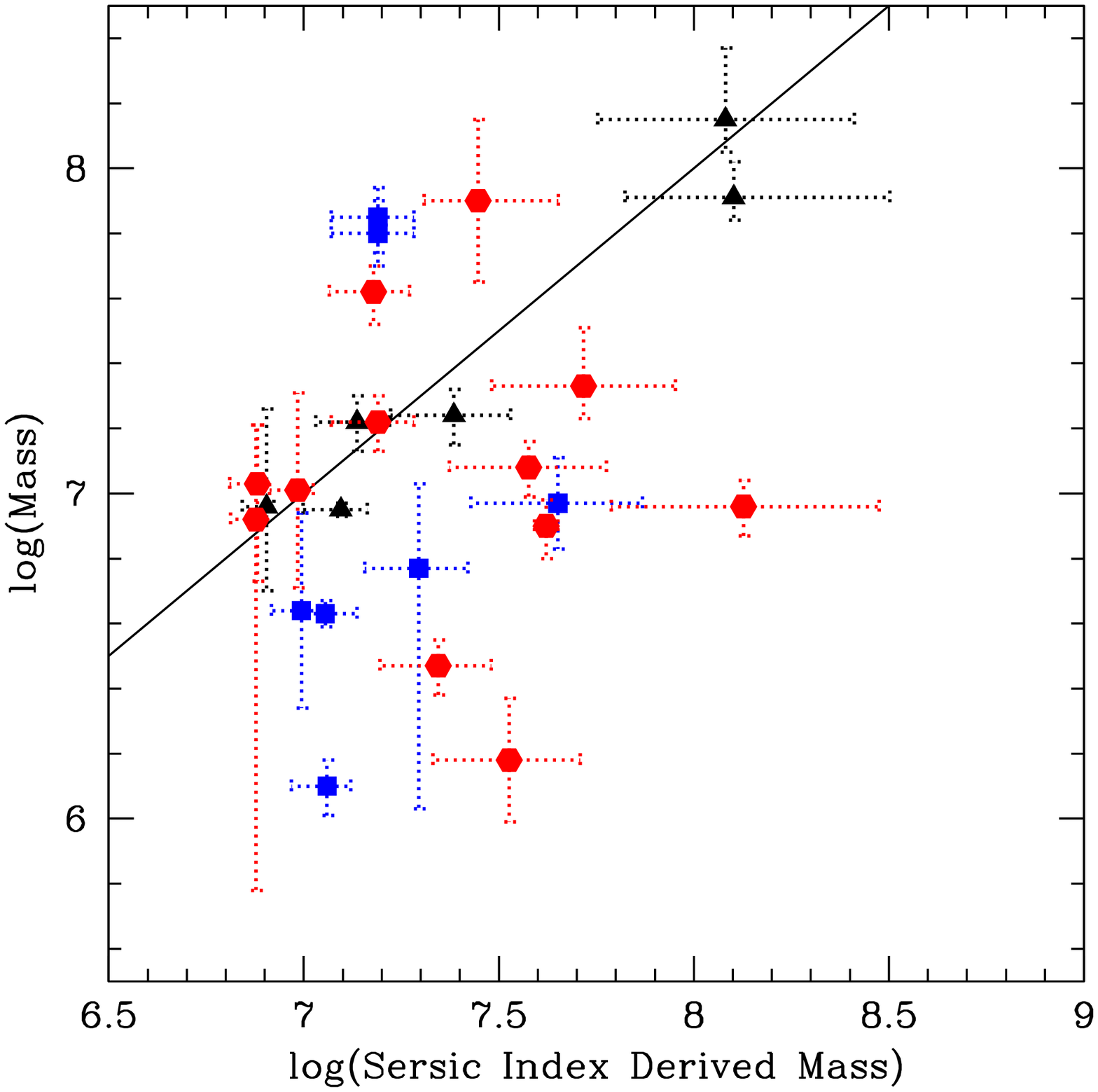}
\caption{\underline{Left}:  S\'ersic index  compared to  pitch angle  for a
  subsample  of  galaxies used  in  this  work.   The black  triangles
  represent  regular (non-barred) spiral  galaxies.  Red  hexagons are
  galaxies with  morphological classifications as  weak bars. Finally,
  blue  squares are  barred galaxies.   \underline{Right}: a  comparison of
  log(Mass)  from the literature  with the  log(S\'ersic-index-derived
  mass) using the relationship  of \citet{Graham2007}.  The solid line
  through  the data  represents  the 1:1  line.  Note that  non-barred
  galaxies  have much  less scatter  about  the 1:1  line than  either
  variety of barred galaxy. The  scatter for the total sample is $\sim
  0.60$ dex, but is only  $0.15$ dex for spiral galaxies without bars,
  $0.73$ and  $0.72$ dex  for barred and  weakly barred  galaxies.  No
  limits are  used in this right  panel, hence there  are fewer points
  than in the panel on the left.}
\label{fig:sersic}
\end{figure*}

The left panel of Figure~\ref{fig:sersic}
%
%
%
%
illustrates the relationship
between  S\'ersic index  and  the  pitch angles  for  galaxies in  our
sample. In the figure, the  black triangles are spiral galaxies without
bars,  red hexagons  are  spiral  galaxies with  weak  bars, and  blue
squares are barred  spirals.  As can be seen from  the figure, there is
little apparent correlation between these  two properties.  This is in
contrast to the fairly  tight relation between S\'ersic-index-produced
masses compared to the masses from the literature, see the right panel
of  Figure~\ref{fig:sersic}.   Yet,  both relations  seem  to  provide
reasonable estimates of SMBH mass.

The right  panel of  Figure~\ref{fig:sersic} compares the  masses from the
literature with the  masses derived from the S\'ersic  index. The same
point types  are used  in the  right panel  as the  left.  A  point of
interest  seems to  be the  low  scatter about  the 1:1  line for  the
regular galaxies  when compared  to barred galaxies,  but that  is not
surprising  since  it is  more  difficult  to  measure $n$  in  barred
galaxies using  our isophotal fitting  technique.  The scatter  for the
total sample is $\sim 0.60$ dex, but is strikingly only $0.15$ dex for
spiral  galaxies  without  bars.   This would  seem  to  suggest  that
S\'ersic index is an extremely  useful tool in estimating $M_{\rm BH}$
for elliptical  galaxies and even non-barred  early-type spirals.  For
later-type  spirals one  has to  be  extremely careful  and adopt  the
approach  of  using  a multi-component  two-dimensional  morphological
fitting routine  (such as  GALFIT; \citealt{Peng2002}).  This  can become
computationally expensive,  especially when  compared to  our approach
for determining spiral pitch angles.

The results of this work suggest that for a grand design spiral galaxy
the $M$--$P$ relation, with its low  scatter compared to other methods when
applied  to  spiral  galaxies,  could   be  the  preferred  method  of
estimating the central  black hole mass. This may even  prove true for
all classes  of spiral  galaxies.  Even  if spectroscopic  methods are
preferred in some cases, a considerable advantage remains that the $M$--$P$
relation requires  only imaging data,  which is becoming  plentiful at
high quality.

Accurate  S\'ersic indices  still  require more  image processing  and
initial  estimates to  determine,  even  when a  code  such as  GALFIT
\citep{Peng2002} or  BUDDA \citep {BUDDA}  is used for  barred spiral
galaxies.  In  these cases,  there are a  larger number  of parameters
that need  to be fit,  three parameters for  the bulge, three  for the
bar, and two  for the disk.  A large amount  of degeneracy may result,
making it more complex to get an accurate handle on the S\'ersic index
than  to measure  the pitch  angle. These  methods have  been used  to
measure large numbers of galaxies \citep[e.g.][]{Gadotti2009}, but they
still  require significant  time  for image  processing. As  discussed
above, unresolved nuclear flux  will create difficulties in determining
S\'ersic index.   Automated codes may have  difficulties disentangling
the unresolved nucleus and the bulge  for distant galaxies, which is an
issue   for  several   galaxies  in   our  sample.    The  method   of
\citet{Davis2012} requires  little in the  way of image  processing in
order to  measure the  pitch angle  of the  spiral arms,  and requires
little  user input.  In  the case  of  late-type non-barred  galaxies,
\citet{Davis2013} compare  SMBH masses (derived from  spiral arm pitch
angles using  the relation here)  with S\'ersic indices  measured from
\citet{Graham2007}, and an encouraging  result is found.  As discussed
in \citet{Davis2013},  there are  strong reasons  to prefer  the pitch-angle-derived  mass  function  to the  S\'ersic-index-derived  mass
function in the  case of late-type galaxies. Nevertheless,  it is very
encouraging that  the evidence  presented here suggests  that S\'ersic
index and pitch angle estimates of  black hole mass are compatible for
non-barred  spirals.   This  raises  the  immediate  prospect  that  a
combination of  these two  approaches (i.e.,   using S\'eric  index for
ellipticals and pitch  angles for spirals) could produce  a black hole
mass  function   for  all  types   of  galaxies  using   imaging  data
alone.  Additionally, a  potential  method for  rapid automated  pitch
angle measurements in development is discussed in \citet{Davis&Hayes}.

This comparison  provides another opportunity to  examine our results.
By comparing the residuals, derived  by subtracting these results from
the masses  in the literature, we  may elucidate whether  the correlations
are independent.  This  will also inform us as to  whether the scatter
can  be reduced  through  a combination  of  several parameters.   For
residuals  derived from  the linear  S\'ersic index  mass relation  of
\citet{Graham2007}   versus  pitch   angle  we   find  a   correlation
coefficient of $-0.38$ with $80.4\%$  significance. This is a moderate
anti-correlation, but essentially  insignificant.  Using the preferred
quadratic   relation   from    \citet{Graham2007},   the   correlation
coefficient  becomes  $0.33$  (weak  correlation)  which  is  $97.3\%$
significant.  This  is a  greater than  $2\sigma$ result.   For pitch
angle  residuals  versus  S\'ersic  index,   we  find  a  Pearson  rank
correlation coefficient  of $0.74$, with a  significance of $99.99\%$,
about a $4\sigma$ result.

\section{Discussion}
\label{sec:Discussion}

\subsection{Comparisons to Other Methods}
\label{sec:discomp}

In considering the fits to the various categories of data (stellar and
gas dynamics,  masers, reverberation mapping, $\sigma$),  it is notable
that the  scatter is less  than $0.48$ dex in  all of the  samples and
subsamples considered.  In  the case of the  direct measurements using
stellar  or  gas  dynamics,  the  scatter  is  $0.39$  dex.   This  is
comparable in scatter to the best of the other galactic features which
are  known to  correlate to  black hole  mass.  In  particular, it  is
comparable to the scatter for the $M$--$\sigma$ relation, $0.44$ dex, for
the full sample of all  galaxies in \citet{Gultekin2009}.  The results
of \citet{Haring2004} produce $\sim 0.3$ dex scatter using a sample of
$30$ galaxies.  These results  rely primarily  on early-type galaxies
(Ellipticals +  S0) to generate  these results,  as only five  of these
galaxies are  spiral galaxies.  This provides  us with  another useful
tool, as bulge luminosity is another imaging-based observable relation
that may be complimentary  in a full census of SMBH  mass.  It is true
that  our correlation  holds  true  only for  one  category of  galaxy
(spiral galaxies),  but many  of the  other correlations  hold greater
uncertainties and scatter  for just this type, which  is a significant
portion  of the  total population  of galaxies.   It is  further worth
noting  that the  scatter for  $M$--$\sigma$, using  only late types,  is
about    $0.56$    dex,    somewhat    worse    than    our    scatter
\citep{Gultekin2009}.  Bulge  luminosity and  S\'ersic index  (both of
which  depend on  measurements  of  the central  core  of the  galaxy)
encounter difficulties with spirals,  especially barred spirals, where
one has to subtract the disk and bar components to find the true bulge
component. Routines such as GALFIT and BUDDA provide powerful tools to
make  the  necessary  measurements   of  the  luminosity  from  bulges
\citep{Peng2002, BUDDA}. These results may be used with relations such
as those  from \citet{Haring2004}  and \citet{Graham2007}  to estimate
SMBH masses.  As one can imagine, the more free parameters you have to
fit  to the  observed surface  brightness profiles  (and there  are at
least five  in the case of  non-barred spiral galaxies, and  as many as
eight free  parameters in  barred spirals),  the greater  the degeneracy
between  these parameters.   As a  result, measuring  the value  of the
S\'ersic  index   determined  from  such  fits   becomes  increasingly
complicated.  In  complex systems  (i.e., those  with bars  and disks,
compared   to   pure  bulge   or   elliptical   galaxies),  SMBH   mass
determinations based  on pitch angle  also appear to be  consistent in
galaxies  with pseudobulges,  where  other relations  appear to  break
down.  It is certainly true that many of the black hole mass functions
published    to     date    concentrate    first     on    early types
\citep[e.g.][]{Marconi2004}. We  intend to use the  $M_{\rm BH}$--pitch
angle relation to redress this imbalance \citep{Davis2013}.
Since spirals will typically have undergone few major mergers in their
history in  comparison with  ellipticals, this technique,  which works
well  for   spirals,  can  help  develop   information  which  applies
particularly to  the accretion history  of black holes, as  opposed to
the merger history.

The  fact  that  various  macroscopic and  morphological  features  of
galaxies correlate  to each other  has been noted since  galaxies were
first  observed  in   any  number  by  Edwin  Hubble   in  the  1920s.
Theoretically, the  possibility that the  black hole at  each galaxy's
center should correlate with these features is by no means certain but
not implausible.   A common assertion based  upon strong observational
evidence is that the mass of  the central black hole correlates to the
mass    of    the    galaxy's    central   bulge    or    core    area
\citep[e.g.,][]{Magorrian1998,          Marconi2003,         Haring2004,
  Kormendy2011A}. The  reason for such  a correlation is  not settled,
though various proposals  have been made. One  such proposed mechanism
assumes that the depth of the potential in the central region in which
the  black hole  resides  governs  the amount  of  available fuel  for
accretion.   If the  black hole  becomes  too large  and accretes  too
rapidly,  radiation pressure  will force  fuel out  of the  well, thus
starving  itself of  further growth.   Thus, the  mass of  the central
region  places   an  upper   limit  on  growth   of  the   black  hole
\citep{Silk&Rees1998}. As long as the  SMBH reaches this limit at some
point during  its life, its mass  should correlate to the  mass of its
bulge. Recently,  it has  been suggested that  the causal  relation is
reversed.  Instead of  the mass of the core controlling  the growth of
the black hole,  it is the mass  of the black hole  which controls the
growth of the core \citep{Jahnke2009}.   Either way, the data suggest
that such a correlation exists in some form.

Another observable which seems to correlate well with SMBH mass is the
S\'ersic  index   of  a  galaxy   ($n$),  as  discussed  above   and  in
\citet{Graham2001,  Graham2007}.   The  S\'ersic  index  of  a  galaxy
measures the light concentration of  a galactic bulge, and thus should
correlate  to the  mass  concentration  of a  galaxy.   As  we saw  in
Section~\ref{sub:sersicres}, this  relation has  potential difficulties
when  dealing  with  a  large  portion of  the  population  of  spiral
galaxies.  This may  be a result of the techniques  adopted to measure
S\'ersic  index in  complicated systems,  but this  alone may  make it
difficult  to  automate the  measurement  of  S\'ersic index,  whereas
techniques  to automate  the  measurement of  spiral  pitch angle  are
already  being  explored  \citep{Davis&Hayes}.  The  relation  between
S\'ersic index and SMBH mass  is however a complimentary relation that
could be  used to estimate  masses for  elliptical and S0  galaxies as
well as confirm results on non-barred bulge-dominated spirals.

Since direct measures of black hole masses are intrinsically difficult
measurements to make, there is great interest in the use of quantities
like bulge luminosity, $\sigma$, and $n$ as markers for the study of a SMBH mass function and its evolution.   There is a
level of discomfort  with the use of such markers  that centers around
the vast  difference between the  scale of  the black hole  (which, in
terms of the  region of strongly curved spacetime, is  similar in size
to our solar system, and in terms of the region within which the black
hole mass dominates the local stellar  mass is only a kiloparsec or so)
and the scale of the galaxy's central core (on the order of $10$ kpc).
Is it  really plausible that reliable  correlations between quantities
on  these greatly  disparate  scales exist?   How  much more  cautious
should we  be in comparing the  same black hole with  a quantity whose
scale  spans  the entirety  of  a  disk galaxy,  a  scale  of tens  of
kiloparsecs?

Nevertheless,  there  are   excellent  observational  and  theoretical
grounds  for believing  that the  pitch  angle of  spiral arms  should
correlate well with the mass of the galaxy's central bulge despite the
significant  difference  in   scale  \citep{Lin&Shu1964,  Bertin1989A,
  Bertin1989B,    Fuchs1991, Fuchs2000, Block1999,    Seigar2004,
  Seigar2005,  Seigar2006,  SKKL08}.   Indeed, not  only  should  this
relation be causal,  but also we can expect it, on  theoretical grounds, to
be quite tight.


\section{Conclusions} \label{sec:Conclusions}

We have  measured spiral arm  pitch angles for  a sample of  $67$ disk
galaxies with previously determined SMBH  masses.  The SMBH masses for
these galaxies were  taken from a variety of  sources which determined
the mass via several methods  including (1) direct measurements using
stellar  or  gas  dynamics,  (2)  maser  modeling,  (3)  reverberation
mapping, (4) stellar velocity dispersion and the $M_{\rm BH}$--$\sigma$
relation,  and (5)  Eddington  limits  for lower  mass  limits on  the
SMBHs. The  results of several fits  to these samples may  be found in
Table~\ref{table:fits}.
\begin{deluxetable*}{lcccc}
\tabletypesize{\scriptsize}
\tablecaption{Fitting Results}
\tablehead{
\colhead{Sample} & \colhead{ Fit } & \colhead{$\chi^2$} & \colhead{Scatter (dex) } & \colhead{Sample Size} 
}
\startdata
 Stellar/Gas, Masers, and Reverberation Mapping\tablenotemark{a}  & $\log(M_{\rm BH}/M_{\odot})  = (8.21\pm0.16)-(0.062\pm0.009)P$   & $ 4.7$ & $0.38$ & $34$ \\    
 Stellar and Gas Dynamics                        & $\log(M_{\rm BH}/M_{\odot})  = (8.66\pm0.43)-(0.088\pm0.031)P$   & $ 1.2$ & $0.39$ & $10$ \\ 
 Masers                                          & $\log(M_{\rm BH}/M_{\odot})  = (7.43\pm0.23)-(0.027\pm0.012)P$   & $ 0.9$ & $0.30$ & $14$ \\  
 Reverberation Mapping                           & $\log(M_{\rm BH}/M_{\odot})  = (8.48\pm0.15)-(0.072\pm0.009)P$   & $ 0.9$ & $0.28$ & $14$ \\ 
 Masers and Reverberation Mapping                & $\log(M_{\rm BH}/M_{\odot})  = (8.09\pm0.45)-(0.058\pm0.025)P$   & $ 3.7$ & $0.38$ & $28$ \\
 $M$--$\sigma$                                      & $\log(M_{\rm BH}/M_{\odot})  = (8.47\pm0.24)-(0.089\pm0.013)P$   & $ 4.9$ & $0.48$ & $23$ \\ 
 All                                             & $\log(M_{\rm BH}/M_{\odot})  = (8.36\pm0.15)-(0.076\pm0.008)P$   & $10.4$ & $0.45$ & $54$ \\  
\enddata
\tablecomments{Column 1: type of galaxy mass measurements used in the sample. Column 2: sample fit. Column 3: $\chi^2$ for the fit. Column 4: scatter about the fit in dex. Column 5: number of galaxies in the sample.}
\tablenotetext{a}{This fit represents our preferred fit to the data.}
\label{table:fits}
\end{deluxetable*}
Our main conclusions are as follows.

\begin{itemize}

\item Using  only galaxies  with direct  SMBH mass  measurements based
  upon stellar  and gas dynamics in  normal galaxies as well  as maser
  modeling and  reverberation mapping in  active galaxies, we  find a
  SMBH mass--pitch angle relation of
\begin{equation}
\log(M_{\rm BH}/M_{\odot}) = (8.21\pm0.16) - (0.062\pm0.009)P.
\nonumber
\end{equation}

\item If  we include  also select indirect  black hole  mass estimates
  which   were   used  in   the   previous   correlation  studied   in
  \citet{SKKL08}, then we find
\begin{equation}
\log(M_{\rm BH}/M_{\odot})  = (8.36\pm0.15)-(0.076\pm0.008)P.  
\nonumber
\end{equation}
This  relation  is statistically  consistent  with  that presented  in
\citet{SKKL08} and virtually identical in slope.

\item  Our scatter  is  comparable to,  or better  than,  that in  the
  $M$--$\sigma$ relation.

\item  Our  technique  does   not  require  observationally  expensive
  spectra.   The  method  is also  cosmologically  independent,  since
  logarithmic spirals are self similar.

\item Using  the relationship of  \citet{Graham2007} with a  sample of
  our  galaxies illustrates  that there  is more  scatter in  Graham's
  relationship for barred galaxies. Meanwhile, our relationship should
  not be  affected by bars.  Thus, two  SMBH mass estimators  which can
  take advantage of imaging data only (pitch angle and S\'ersic index)
  are complementary.

\end{itemize}

Equation~\ref{eqn:final}  is  extremely   useful  for  measuring  SMBH
masses.   In  \citet{Davis2012}, we  showed  that  pitch angle  can  be
measured  extremely  reliably.  In  future  papers,  we intend  to  use
Equation~\ref{eqn:final} to  determine a local SMBH  mass function for
spiral galaxies \citep{Davis2013}, and for higher-$z$ spirals, provided
we  can determine  how  (or  if) the  SMBH  mass--pitch angle  relation
evolves as  a function of look-back  time. We can often  measure pitch
angle  to  within  a  relative  error  of  $3^\circ$  or  less.   This
translates to a relative error in the logarithm of the black hole mass
of $4\%$.

\acknowledgments 

The authors  gratefully acknowledge  support for  this work  from NASA
Grant NNX08AW03A  and NSF REU  Site Grant 1157002.  This  research has
made  use  of the  NASA/IPAC  Extragalactic  Database (NED)  which  is
operated  by the  Jet Propulsion  Laboratory, California  Institute of
Technology,  under contract  with the  National Aeronautics  and Space
Administration.  The  authors also thank Heather  Berrier for
helpful conversations  and suggestions.  We are  especially grateful to
Wayne Hayes and Darren Davis of the University of California at Irvine
for measuring a number of our galaxies with their pitch angle code for
confirmation purposes  and for many helpful  discussions on measurement
issues.


\appendix

\section{Spiral Density Waves and Pitch Angle}
\label{sec:DensityWaves}

In the modal density wave theory it  is fairly straightforward to  see that
the pitch angle  of the spiral arms must vary  inversely with the mass
of the central bulge of  the galaxy.
From \citet{Bertin&Lin1996} we have
\begin{equation}
\tan i = {m \over r k},
\end{equation}
where $i$ is the pitch angle of  the spiral pattern, $m$ is the number
of spiral arms, $r$ is the radial position in the disk, and $k$ is the
``local radial wavenumber'' which is not constant over  the disk but
is related to the local wavelength  of the density waves by $\lambda=2
\pi/k$.

\citet{Shu1984} deals with
the case where  the pitch  angle is relatively small,  which is
  true for Saturn but also for  the tightest spiral arms in galaxies,
and in the treatment found there one sees how
\begin{equation}
|k| = {D \over 2 \pi G \sigma_o},
\end{equation}
where $\sigma_o$ is the surface mass density in the disk and $D$ is an
expression  which can  be  understood as  the  distance, in  frequency
terms, from Lindblad resonance of the gravitational potential.

A Lindblad resonance occurs when
\begin{equation}
\omega - m \Omega = \pm \kappa,
\end{equation}
where  $\Omega$  is the  $\theta$  (or  tangential) frequency  of  the
gravitational potential of the central mass of the galaxy (for a point
mass this  is $\Omega=\sqrt{G M/ r^3}$,  where $M$ is the  mass of the
planet  and $r$  is the  radial distance  from it)  and  $\kappa$ is  the
epicyclic frequency or  radial frequency of the same  potential. For a
point  mass this  is  the same  as $\Omega$.   Finally,  $\omega$ is  a
frequency associated with the particle  orbiting in this potential, in
practice likely to be some  kind of forcing frequency which encourages
the  particle to  move with  a frequency  not quite  that of  the main
potential.   For instance,  in  the  case of  spiral  density waves  in
Saturn's rings this  $\omega$ would be associated  with the perturbing
influence of  a nearby  moon.  Since both  $\Omega$ and  $\kappa$ will
vary with  radial coordinate $r$,  it follows  that there will  only be
certain values of $r$ for which  the above relation is satisfied. Such
locations are called Lindblad  resonances and denoted $r_{\rm{L}}$. In
the  theory,   spiral  density   waves  emanate  from   such  Lindblad
resonances.

The definition of $D$ is
\begin{equation}
D = \kappa^2 - (\omega - m \Omega)^2
\end{equation}
and clearly at a Lindblad resonance  it follows that $D=0$.  As a wave
moves  away  from this  resonance,   $D$  will no  longer  be
zero. It is easy  to show, by a Taylor series  expansion, that the key
to the pattern of density waves  propagating from the resonance is the
first derivative of $D$.

The details of calculating $D$  and its derivative are complex because
we are dealing with  a system which does not encourage  the use of the
usual simplifying  assumptions. The two main  frequencies $\Omega$ and
$\kappa$ are  generated by a  decidedly non-point source,  the central
region of the  galaxy, and there are many perturbing  effects from all
of the stars and other material in the disk. Nevertheless, it is 
obvious that $D \propto M$ where $M$ is the mass of the
central gravitational source (it would be the mass of Saturn in the case
of Saturn's rings).

Recall that  if we have  a point mass source, then  we would have  $\Omega =
\sqrt{G M/r^3}$.   In practice, this  is far from  the case, but  we do
expect that $\Omega = \sqrt{G M} f(r)$ where the particular functional
dependence on radial distance is uncertain,  but where $M$ is known to
be all of the  mass inside the orbit of the  particular star or object
whose motion we are following.  Especially for stars in the inner part
of  the disk,  more specifically  for density  waves emitted  from the
inner Lindblad  resonance, this $M$ will  be close to the  mass of the
central bulge of the disk galaxy.

We must also consider the radial  frequency of the orbit (which is, in
general, eccentric) and we recall that  for an orbit in the equatorial
plane of an axisymmetric potential
\begin{equation}
\kappa^2= {1\over r^3} {d\over dr}\big[(r^2\Omega)^2\big]
={2\Omega \over r} {d\over dr} \big(r^2 \Omega\Big).
\end{equation}
In the point  mass case $\kappa$=$\Omega$, but
if $\Omega$ has the generic form given above then we find that
\begin{equation}
\kappa^2={2\sqrt{G M} f(r) \over r} \sqrt{G M}(2 r f(r) + r^2
df(r)/dr) \\=2 G M \Big(2 f^2 + r f {df\over dr}\Big).
\end{equation}
Thus provided  $\Omega$ is  proportional to  $\sqrt{M}$, we  find that
$\kappa$ will also be proportional to $\sqrt{M}$. The frequency $\omega$ is not so constrained in
general, but near the Lindblad resonance  it must be close to $\Omega$
and $\kappa$ and  so it will  also be  effectively proportional to
$\sqrt{M}$.   Since  our  density  waves  originate  at  the  Lindblad
resonances, it follows that this is true in our case.

We return to the key quantity $D$ and begin from
\begin{equation}
D = \kappa^2 -  (\omega - m \Omega)^2 = 2 G M  \Big(2 f^2+r f {df\over
  dr}\Big) -\Big(\omega - m \sqrt{G M} f\Big)^2.
\end{equation}
Since near resonance we expect $\omega$  to have a value such that the
second term on the right is small, we rewrite it as
\begin{equation}
D = 2 G M (2 f^2+r f {df\over dr}) - G M f^2\Big({\omega\over \sqrt{G M} f}
-m\Big)^2.
\end{equation}
So we find that $D \propto M$,  and  the derivative of
$D$ with  respect to $r$  will also be  proportional to $M$,  which is
(approximately)  constant,  especially  to  small  changes  in  radial
distance.  The fact that $D \propto M$ means that
\begin{equation}
\tan  i \propto\sigma_o/M.
\label{eqn:string}
\end{equation}

\bibliography{ms}

\end{document}